\documentstyle[12pt]{article}                  
\input epsf
\makeatletter
\def\nothing#1{}
\newdimen\earraycolsep
\renewcommand{\theequation}{\arabic{section}.\arabic{equation}}

\def\author#1{{\pretolerance=10000 \raggedright \advance \leftskip by 1in \noindent #1 \vskip 1pc}}
\def\affiliation#1{{\advance\leftskip by 1in \noindent #1 \vskip -1pc}}

\def\AmS{{\protect\the\textfont2%
        A\kern-.1667em\lower.5ex\hbox{M}\kern-.125emS}}

\def\p@LaTeX{{\family{times}\series{m}\shape{n}\selectfont L\kern-.36em\raise.3ex\hbox{\scriptsize A}\kern-.15em T\kern-.1667em\lower.7ex\hbox{E}\kern-.125emX}}
\newlength{\colwidth}

\def\@oddhead{\hfil}
\def\@evenhead{\hfil}
\long\def\@makecaption#1#2{\vskip 10\p@ {#1 #2\par}}
\def\centerfig#1#2#3#4{\vspace*{#2}\relax\centerline{\hbox to#1{\special{#4:#3.#4 x=#1, y=#2}\hfil}}}
\newbox\@atbox
\def\@nbibitem#1{\noindent \hangindent=2pc \hangafter=1
\immediate\write\@auxout{\string\bibcite{#1}{\arabic{enumi}}}}
\def\numbibliography{%
\def\newblock{\hskip .11em plus.33em minus.07em}%
\let\bibitem\@nbibitem}
\def\endnumbibliography{\par\egroup}
\makeatother

\def \D {\Delta_{\A}}

\def \Dir {{\cal D}}

\def \Triangle {\bigtriangleup}
\def \ra {\rangle}
\def \la {\langle}

\def \be {\begin{equation}}
\def \eq {\end{equation}}
\def \bee{\begin{eqnarray}}
\def \eqq{\end{eqnarray}}
\def \nn {\nonumber}

\def \a {\alpha}
\def \b {\beta}
\def \g {\gamma}
\def \d {\delta}
\def \l {\lambda}
\def \r {\rho}

\def \zb {\bar{z}}

\def \wb{\bar{w}}
\def \del{\partial}
\def \delb{\bar{\partial}}
\def \deltb {\bar{\delta}}
\def \Zp {{\cal Z_+}}
\def \Zm {{\cal Z_-}}
\def \H {{\cal  H}}

\def \O  {{\cal O}}

\def \V {{\cal V}}
\def \W {{\cal W}}
\def \U {{\cal U}}
\def \CR {{\cal R}}

\def \Db{\bar{D}}

\def \etb {\bar{\eta}}
\def \xb {\bar{x}}
\def \yb {\bar{y}}
\def \xib {\bar{\xi}}

\def \CC{{\cal C }}
\def \CA{{\cal A }}

\def \D{{\cal D }}

\def \O{{\cal O }}

\def \I {{\cal I}}
\def \R {\hat{R}}
\def \RR {\tilde{R}}
\def \Pt {\tilde{\Phi}}
\def \Md {M^\dagger}

\def \ab {{\bar{a}}}
\def \bb {{\bar{b}}}
\def \cb {{\bar{c}}}
\def \db {{\bar{d}}}
\def \r {\rho}
\def \om {\omega}

\def \s {\;\;}

\def\Bbb#1{{#1\kern-.647em #1}}

\def \s{\sigma}
\def \eps{\epsilon}
\def \om{\omega}

\def \be{\begin{equation}}
\def \eq{\end{equation}}
\def \ba*{\begin{eqnarray*}}
\def \ea*{\end{eqnarray*}}
\def\tens{\mathop{\otimes}}

\def \Real{{\bf R}}
\def\Z{{\bf Z}}

\def\C{{\bf C}}

\newtheorem{cn}{Construction}

\begin{document}


\begin{center}
\hfill    LBL-39275 \\ 
\hfill    UCB-PTH-96/38 \\ 
\vskip .2in
{\Large \bf SOME COMPLEX QUANTUM MANIFOLDS \\
\vskip .1in
AND THEIR GEOMETRY}
\footnote{Lectures presented at the meeting on
``Quantum Fields and Quantum Space Time'',
NATO Advanced Study Institute, Institut D'Etudes Scientifiques De Cargese,
July 1996.}
\footnote{This work was supported in part by the Director, Office of Energy
Research, Office of High Energy and Nuclear Physics, Division of High
Energy Physics of the U.S. Department of Energy under Contract
DE-AC03-76SF00098 and in part by the National Science Foundation under
grant PHY-9514797.}
\vskip .3in
\end{center}

\author{Chong-Sun Chu
\footnote{e-mail address: cschu@physics.berkeley.edu}, 
Pei-Ming Ho
\footnote{e-mail address: pmho@physics.berkeley.edu} 
and Bruno Zumino}  

\affiliation{
\begin{center}
Theoretical Physics Group\\
Ernest Orlando Lawrence Berkeley National Laboratory\\
University of California, Berkeley, California  94720\\
and\\
Department of Physics\\
University of California, Berkeley, California  94720
\end{center}
}

\begin{abstract}
After recalling briefly some basic properties of the quantum group $GL_q(2)$,
we study the quantum sphere $S_q^2$, quantum projective space $CP_q(N)$
and quantum Grassmannians as examples of complex (K\"{a}hler) quantum
manifolds.
The differential and integral calculus on these manifolds are discussed.
It is shown that many relations of classical projective geometry
generalize to the quantum case.
For the case of the quantum sphere a comparison is made with A. Connes' method.
\end{abstract}

\tableofcontents

\section{INTRODUCTION}        
\setcounter{equation}{0}

Quantum spheres can be defined in any number of dimensions by
normalizing a vector of quantum Euclidean space \cite{FRT}.
The differential calculus on quantum Euclidean space \cite{oz}
induces a calculus on the quantum sphere.
The case of two-spheres in three-space is special in that there are
many more possibilities.
These have been studied by P. Podle\'{s} \cite{P1,P2,P3,P4} who has defined
quantum spheres as quantum spaces on which quantum $SU_q(2)$ coacts.
He has also developed a noncommutative differential calculus on them.
In these lectures we consider, following \cite{CHZS2}, a special case of Podle\'{s} spheres which is
one of those special to three space dimensions.
In this case the quantum sphere $S_q^2$ is the analogue of the classical sphere
defined as $SU(2)/U(1)$ or as isomorphic to $CP(1)$.
We also define a stereographic projection and describe the coaction of $SU_q(2)$
on the sphere by fractional transformations on the complex variable in the plane
analogous to the classical ones.
The quantum sphere appears then as the quantum deformation of the classical
two-sphere described as a complex K\"{a}hler manifold.
We discuss the differential and integral calculus on $S_q^2$ and the 
action of $SU_q(2)$ vector fields on it.
Finally, following \cite{CHZBS2},
 we show that one can define on braided copies of $S_q^2$ invariant
anharmonic cross ratios analogous to the classical ones.
All this is done in Sec.\ref{sq2-alg},
after recalling briefly in Sec.\ref{GL2} the basic properties of $GL_q(2)$ and $SU_q(2)$.

The above results are generalized in Secs.\ref{CPqN} and 5
to quantum $CP_q(N)$ and in Sec.\ref{Grass} to
quantum Grassmannians \cite{CHZCPN}.
These quantum spaces appear as complex K\"{a}hler quantum manifolds
which can be described in terms of homogeneous or inhomogeneous coordinates.
Differential and integral calculus can be defined on them as well as
the quantum analogues of projective invariants. For the general case of Grassmannians,
we do not give explicit formulas for the integral and for the projective invariants. 
They should not be hard to derive by analogy with the $CP_q(N)$ case.

The type of quantization described here has the property that there exists
a special quantum (connection) one-form which generates the differential calculus
by taking commutators or anticommutators of it with functions or forms
(see Eq.(\ref{xi-eta}) below).
This one-form is closely related to the K\"{a}hler form which can be
obtained from it by differentiation.
In the Poisson limit our quantization gives Poisson
brackets not only between functions but also between functions and forms and
between forms. The special one-form generates the calculus
by taking Poisson brackets with functions or forms and the K\"{a}hler
form can still be obtained from it by differentiation.
Our Poisson structure on the manifold is singular and is not the standard one
which is obtained by taking the K\"{a}hler form as symplectic form.
Nevertheless, our Poisson structure is intimately related to the K\"{a}hler
form, as just explained. For the algebra of functions on the sphere,
this singular Poisson structure 
was already considered in \cite{We}.

All formulas and derivations of Sec.\ref{sq2-alg} can be easily modified,
with a few changes of signs, to describe the quantum unit disk and the coaction
of quantum $SU_q(1,1)$ on it, as well as the corresponding invariant
anharmonic ratios.
This provides a quantum deformation of the Bolyai-Lobachevski\v{i} non-Euclidean
plane and of the differential and integral calculus on it.
The modified equations can be guessed very easily and will not be given here.
It should be mentioned that the commutation relations between $z$ and $\zb$
for the unit disk are consistent with a representation of $z$ and $\zb$ as
bounded operators in a Hilbert space.
This is to be contrasted with the case of the quantum sphere where $z$ and $\zb$
must be unbounded operators.
The developments of Sec.\ref{CPqN} and \ref{Grass} can similarly be modified,
again with some changes of signs, to describe a quantum deformation of various
higher dimensional non-Euclidean geometries.

Finally, in Appendix A, we try to re-formulate the differential and integral
calculus on the quantum sphere in a way as close as possible to Connes'
formulation of quantum Riemannian geometry \cite{Con}.

We will use the following notations throughout the paper:
\be
   \l=q-q^{-1}
\eq
and
\be [n]_q =\frac{q^{2n}-1}{q^2-1}.
\eq

\section{$GL_q(2)$ AS EXAMPLE OF A QUANTUM GROUP}\label{GL2}
\setcounter{equation}{0}

The simplest example of a matrix quantum group \cite{W1} as
a Hopf algebra \cite{Swe} is $GL_q(2)$ \cite{FRT}.
It is a one-parameter deformation of the classical group $GL(2)$.
The algebra of functions on $GL(2)$
is generated by the elements $\a, \b, \g, \d$
in the fundamental representation
\be
   \pi(g) = \left(
                \begin{array}{ll}
                   \a(g) & \b(g) \\
                   \g(g) & \d(g)
                \end{array}
                \right)
\eq
for $g\in GL(2)$.
The algebra $\CA$ of functions on $GL_q(2)$ has the following
Hopf algebra structure.
\begin{enumerate}

\item {\em Algebra}:

The multiplication in $\CA$ is noncommutative and the commutation
relations are given compactly as
\be
   \label{RTT}
   \R^{ij}_{kl}T^k_m T^l_n = T^i_k T^j_l\R^{kl}_{mn}
\eq
in terms of the quantum matrix
\be
   T = \left(
                \begin{array}{ll}
                   \a & \b \\
                   \g & \d
                \end{array}
                \right)
\eq
and the $\R$-matrix
\be \label{R-hat-SU2}
   \R = \left(
                \begin{array}{llll}
                   q & 0 & 0 & 0 \\
                   0 & \l & 1 &0 \\
                   0 & 1 & 0 & 0 \\
                   0 & 0 & 0 & q
                \end{array}
                \right),
\eq
where $q$ is a complex number.

The classical limit is obtained
when the deformation parameter $q\rightarrow 1$,
the $\R$-matrix becomes the permutation matrix.

Explicitly the commutation relations are
\bee
   &\a\b=q\b\a, &\a\g=q\g\a, \\
   &\b\d=q\d\b, &\g\d=q\d\g,  \label{GLq2} \\
   &\b\g=\g\b,  &\a\d-\d\a=\l\b\g.
\eqq

The self-consistency of the commutation relations
is guaranteed by the quantum Yang-Baxter relation
\be
   \label{Yang-Baxter}
   \R^{ij}_{i'j'}\R^{j'k}_{j''n}\R^{i'j''}_{lm}
   =\R^{jk}_{j'k'}\R^{ij'}_{lj''}\R^{j''k'}_{mn}.
\eq

\item {\em Coproduct}:

The coproduct of a generator is defined by
the matrix multiplication
\be   
   \left(
   \begin{array}{ll}
      \Delta(\a) & \Delta(\b) \\
      \Delta(\g) & \Delta(\d)
   \end{array}
   \right) =
   \left(
   \begin{array}{ll}
      \a\otimes\a+\b\otimes\g & \a\otimes\b+\b\otimes\d \\
      \g\otimes\d+\d\otimes\g & \g\otimes\b+\d\otimes\d
    \end{array}
    \right).
\eq
This formula is the same as the classical one.
The coproduct 
is a linear map and  an algebra homomorphism.
Another equivalent way to say that the coproduct is an algebra
homomorphism is to say that the algebra is {\em covariant} under
the left transformation
\be
   \label{l-transf}
   T\rightarrow T''=TT'
\eq
or the right transformation
\be
  \label{r-transf}
  T\rightarrow T''=T'T,
\eq
where $T'$ is another quantum matrix satisfying (\ref{RTT}) whose entries
commute with the entries of $T$.
The left- or right-covariance of the algebra means that
the commutation relations among the entries of $T''$
are the same as those for
the corresponding entries of $T$.

\item {\em Counit}:

The definition of the counit on generators also coincides
with the classical case:
\be
   \left(
   \begin{array}{ll}
      \eps(\a) & \eps(\b) \\
      \eps(\g) & \eps(\d)
   \end{array}
   \right) =
   \left(
   \begin{array}{ll}
      1 & 0 \\
      0 & 1
    \end{array}
    \right).
\eq
In short, $\eps(T)=I$, where $I$ is the identity matrix.
For the counit to be an algebra homomorphism,
we need $I$ to be a quantum matrix.
This is true since $I^i_j=\d^i_j$ and the RTT relation (\ref{RTT})
is trivially satisfied.

\item {\em Coinverse}:

The coinverse of the generators is defined so that they form
the inverse  matrix $T^{-1}$. 
It is
\be
   \left(
   \begin{array}{ll}
      S(\a) & S(\b) \\
      S(\g) & S(\d)
   \end{array}
   \right) = (det_q(T))^{-1}
   \left(
   \begin{array}{cc}
      \d & -q^{-1}\b \\
      -q\g & \a
    \end{array}
    \right)=T^{-1},
\eq
where $det_q(T)=\a\d-q\b\g$ is called the quantum determinant of $T$.
The quantum determinant is central in $\CA$
(it commutes with everything in $\CA$) and is assumed not to vanish.
\end{enumerate}

This concludes our brief description of the quantum group $GL_q(2)$
as a Hopf algebra.

Because the quantum determinant is central,
it is consistent with the algebra to impose an additional condition
\be
   \label{det=1}
   det_q(T)=1.
\eq
What we obtain after imposing (\ref{det=1}) is the deformation
of $SL(2)$, naturally named $SL_q(2)$.

A further step can be taken to get $SU_q(2)$.
We define the $*$-involution on $SL_q(2)$ for real $q$ by
\be
   T^{\dagger}=
   \left(
   \begin{array}{ll}
      \a^* & \g^* \\
      \b^* & \d^*
   \end{array}
   \right) =
   \left(
   \begin{array}{cc}
      \d & -q^{-1}\b \\
      -q\g & \a
    \end{array}
    \right)=T^{-1}.
\eq
The commutation relations are covariant under this $*$-involution.

The $*$-involution reverses a product: $(ff')^*=f'^* f^*$
and is complex conjugation on complex numbers.
It corresponds to the Hermitian conjugation
when one realizes the algebra as the algebra of operators
on a Hilbert space.

Everything we mentioned in this section can be generalized
to $GL_q(N)$, $SL_q(N)$ and $SU_q(N)$.
These and the $q$-deformation for other classical groups
are given in \cite{FRT}.

\section{THE COMPLEX QUANTUM MANIFOLD $S^2_q$}\label{sq2-alg}
\setcounter{equation}{0}

In \cite{P1} Podle\'{s} described  a family
of quantum spheres.
They are compact
\footnote{Their classical limit is compact.}
quantum spaces with the quantum symmetry $SU_q(2)$.
That is, the algebra $X$ of functions on the quantum spheres
is covariant under an $SU_q(2)$ transformation.

By studying the representations 
of the universal enveloping algebra of $SU_q(2)$
as in the classical case,
one finds the quantum Clebsch-Gordan coefficients
\cite{Res} which one must use to compose or decompose representations.

A classical sphere can be specified
in terms of Cartesian coordinates
as $x^2+y^2+z^2=r^2$.
The vector
$(e_{+}, e_0, e_{-})=(\frac{1}{\sqrt{2}}(x+iy),z,\frac{1}{\sqrt{2}}(x-iy))$
transforms as a spin-$1$ representation
under $SU(2)$.
In the deformed case
we can use the quantum Clebsch-Gordan coefficients
to find commutation relations
covariant under the linear transformation
of the vector $(e_{+}, e_0, e_{-})$
as a $j=1$ representation of $SU_q(2)$.
It is \cite{P1}
\bee
&e_+ e_- -e_- e_+ +\l e_0^2=\mu e_0, \label{ee1}\\
&qe_0 e_+ -q^{-1}e_+ e_0=\mu e_+, \\
&qe_- e_0 -q^{-1}e_0 e_-=\mu e_-
\eqq
and
\be
e_0^2+qe_- e_+ +q^{-1}e_+ e_- =s, \label{ee4}
\eq
where $\mu\in\Real$ and $s>0$.
We have in addition to $q$ two free parameters $\mu$ and $s$,
where $s$ can always be scaled to a fixed number.
Only $\mu$ labels inequivalent quantum spheres.
With the $*$-involution $q^*=q$,
$e_{+}^*=e_{-}$ and $e_0^*=e_0$,
it gives a $C^*$-algebra.

A particularly interesting case is when
this algebra is equivalent to the quotient $SU_q(2)/U(1)$ \cite{BM1,Eg}.
The classical $U(1)$
\footnote{
Since $U(1)$ is one-dimensional
there are no commutation relations to deform.
}
is represented as a subgroup of $SU_q(2)$ by
\be
   \left(
   \begin{array}{cc}
      U & 0 \\
      0 & U^{-1}
   \end{array}
   \right),
\eq
where $U^*=U^{-1}$.
It can be checked that this is an $SU_q(2)$-matrix.
$SU_q(2)$ transforms under right multiplication by this matrix as
\be
   \left(
   \begin{array}{cc}
      \a & \b \\
      \g & \d
   \end{array}
   \right)\rightarrow
   \left(
   \begin{array}{cc}
      \a & \b \\
      \g & \d
   \end{array}
   \right)
   \left(
   \begin{array}{cc}
      U & 0 \\
      0 & U^{-1}
   \end{array}
   \right),
\eq
which is again an $SU_q(2)$-matrix by taking
$U$ to commute with $\a,\b,\g,\d$.

The algebra of functions $X$ on the quantum sphere
$S_q^2=SU_q(2)/U(1)$
is the subalgebra of the algebra of functions on $SU_q(2)$
which is invariant under this $U(1)$ transformation.
It is generated by $\a \b, \b\g$ and $\g\d$, which can be related to
$e_{+}, e_{-}, e_0$
for $\mu=\l$ and $s=1$ by
\be 
e_0 = 1+q^{-1}[2]_q\b\g, \quad 
e_+ = q^{-3/2}[2]_q^{1/2}\a\b
\eq
and
\be
e_- = -q^{-1/2}[2]_q^{1/2}\g\d.
\eq

One can construct a stereographic projection of the sphere. Define
\bee \label{ze}
z = -q^{1/2} [2]_q^{1/2} e_+ (1-e_0)^{-1} = \a \g^{-1}, \nn\\
\zb = -q^{1/2} [2]_q^{1/2} (1-e_0)^{-1} e_- = -\d \b^{-1},
\eqq
which classically is the projection from the north pole of the sphere to
the plane tangent to the south pole with coordinates $z, \zb$. 
Using (\ref{ze}) and the properties of $SU_q(2)$, one obtains easily
the commutation relation
\be \label {szz}
 z \zb =q^{-2} \zb z +q^{-2}-1,
\eq
or equivalently
\be
   (1+z\zb)=q^{-2}(1+\zb z);
\eq
and the $*$-involution  
\be
z^*=\zb.
\eq

An equivalent description of $S_q^2$ can be obtained from
the $SU_q(2)$ left-covariant complex quantum plane
with coordinates $\{x,\bar{x},y,\bar{y}\}$ satisfying
\bee \label{qplane2}
xy = q yx, & y \yb = \yb y, \nn \\
x \yb = q \yb x, & x \xb = \xb x - q\l \yb y 
\eqq
and their $*$-involutions,
by considering the subalgebra generated by
the inhomogeneous coordinates
\be \label{zx}
z=x y^{-1}, \quad \zb= \yb^{-1} \xb.
\eq

It is easy to obtain the inverse relation of (\ref{ze}). It is
\bee
&e_0 = 1-[2]_q\r^{-1},\nn\\
&e_+ = -q^{-1/2} [2]_q^{1/2}z\r^{-1}, \nn \\ 
&e_- = -q^{-1/2} [2]_q^{1/2}\r^{-1}\zb,
\eqq
where $\r=1+\zb z$.

The $SU_q(2)$ transformation on $SU_q(2)$
induces rotations on the sphere.
In terms of the coordinates $z,\zb$ it is the fractional transformation:
\be
\label{z-transf}
z \rightarrow (a z +b)(c z+d)^{-1}, \quad
\zb \rightarrow -(c-d \zb)(a-b \zb)^{-1},
\eq
where $\pmatrix{a&b\cr c&d} \in SU_q(2)$
and $a,b,c,d$ commute with $z$ and $\zb$.
Eq.(\ref{szz}) is covariant
under this fractional transformation.

We will denote the $*$-algebra generated by $z$ and $\zb$ as $\CC^+$.
Classically $\CC^+$ is the algebra of functions on the plane.
Notice that (\ref{szz}) for this plane differs from the usual 
quantum plane by an additional inhomogeneous constant term. 


\subsection{Differential Calculus} \label{DC-sq2}

In Refs.\cite{P2,P3,P4}, differential structures on $S^2_q$ are studied and classified. 
In this section, we give a differential calculus on the patch $\CC^+$
in terms of the complex coordinates $z$ and $\zb$.
Just as the algebra of functions on $\CC^+$ can be inferred from that
 of $SU_q(2)$,
so can the differential calculus.

For $SU_q(2)$ there are several forms of differential 
calculus\cite{Wcal1,Wcal2,SWZ}: 
the 3D left- or right-covariant differential calculus, 
and the $4D_{+}, 4D_{-}$ bi-covariant calculi.
The 4D bi-covariant calculi have one extra dimension in their space of one-forms compared with the classical case. 
The right-covariant calculus will not give a projection on $\CC^+$ in 
a closed form in terms of $z$, $\zb$, which are defined to transform from the left.
Therefore we shall choose the left-covariant differential calculus.

It is straightforward to obtain the following relations from those for $SU_q(2)$:
\bee
   &z dz=q^{-2}dz z,     & \zb dz=q^{2}dz \zb,  \label{zdz}\\
   &z d\zb=q^{-2}d\zb z, & \zb d\zb=q^{2}d\zb \zb, \\
   &(dz)^{2}=(d\zb)^{2}=0 \label{dzdz}
\eqq
and
\bee
   &dz d\zb=-q^{-2}d\zb dz.
\eqq

We can also define derivatives $\del$, $\delb$ such that on functions
\be
d=dz\del+d\zb\delb.
\eq
From the requirement $d^{2}=0$ and the undeformed Leibniz rule for $d$ together with 
Eqs.(\ref{zdz}) to (\ref{dzdz}) it follows that:
\begin{eqnarray}
\label{delz}
   &\del z=1+q^{-2}z \del, &\del \zb=q^{2}\zb \del,   \\ 
   &\delb z=q^{-2}z \delb, &\delb \zb=1+q^{2}\zb \delb
\eqq
and 
\bee
\label{deldelb}
   &\del \delb=q^{-2}\delb \del. 
\eqq
It can be checked explicitly that these commutation relations are covariant
under the transformation (\ref{z-transf}) and
\bee
   &dz\rightarrow (dz)(q^{-1}cz+d)^{-1}(cz+d)^{-1}, \\
   &\del\rightarrow (cz+d)(q^{-1}cz+d)\del,
\eqq
which follow from (\ref{z-transf}) and the fact that the differential $d$ is invariant.
We hope that there is no confusion: $(dz)$ is the differential of $z$
rather than the quantum group element $d$ times $z$.

The $*$-structure also follows from that of $SU_q(2)$:
\begin{eqnarray}
   &(dz)^{*}=d\zb,  \label{dz*}\\
   &\del^{*}=-q^{-2}\delb+(1+q^{-2})z \rho^{-1},  \label{del*}\\
   &\delb^{*}=-q^{2}\del+(1+q^{2}) \rho^{-1}\zb \label{delb*},
\end{eqnarray}
where the $*$-involution inverts the order of factors in a product.

The inhomogeneous terms on the RHS of the Eqs.(\ref{del*}) and (\ref{delb*}) reflect the fact that
the sphere has curvature.
Incidentally, all the commutation relations in this section admit another possible involution:
\bee 
\label{Cdz*}     &(dz)^{*}=d\zb, \\ 
\label {Cdel*}   &\del^{*}=-q^{2}\delb, \\  
\label{Cdelb*}   &\delb^{*}=-q^{-2}\del. 
\eqq
This involution is not covariant under the fractional transformations and
cannot be used for the sphere.
However, it can be used when we have a quantum plane defined by the same algebra of functions and calculus.
We shall take Eqs.(\ref{zdz}) to (\ref{delb*}) as the definition of the differential calculus on the
patch $\CC^+$.

It is interesting to note that there exist two different types of symmetries in the calculus.
The first symmetry is that if we put a bar on all unbarred variables ($z$, $dz$, $\del$),
take away the bar from any barred ones and at the same time replace $q$ by $1/q$ in any statement about 
the calculus, the statement is still true.

The second symmetry is the consecutive operation of the two $*$-involutions above, so that
\bee
   &\del\rightarrow -q^{2}\delb^{*}=q^{4}\del-q^{2}(1+q^{2}) \rho^{-1}\zb, \\
   &\delb\rightarrow -q^{-2}\del^{*}=q^{-4}\delb-q^{-2}(1+q^{-2})z \rho^{-1},
\eqq 
with $z, \zb, dz, d\zb$  unchanged. 
This replacement can be iterated $n$ times and gives a symmetry which resembles that of a gauge transformation on a line bundle:
\bee
   \del\rightarrow \del^{(n)} &\equiv& q^{4n}\del-q^{2}[2n]_{q} \rho^{-1}\zb \nn \\
                              &=& q^{4n} \rho^{2n}\del \rho^{-2n}, \\
   \delb\rightarrow \delb^{(n)} &\equiv& q^{-4n}\delb-q^{-2}[2n]_{1/q}z \rho^{-1} \nn \\
                                &=& q^{-4n} \rho^{2n}\delb \rho^{-2n}.
\eqq
For example, we have 
\be \del^{(n)} z = 1 + q^{-2} z \del^{(n)}.
\eq 
Making a particular choice of $\del, \delb$ is like fixing a gauge. 

Many of the features of a calculus on a classical complex manifold are preserved.  
Define $\delta=dz\del$ and $\deltb=d\zb\delb$ as the exterior derivatives on
the holomorphic and antiholomorphic functions on $\CC^+$ respectively. We have:
\bee
   &\left[\delta,z\right]=dz, \quad \left[\delta,\zb\right]=0, \\
   &\left[\deltb,z\right]=0, \quad \left[\deltb,\zb\right]=d\zb, \\
   &d=\delta +\deltb.
\eqq
The action of $\delta$ and $\deltb$ can be extended consistently on forms as follows
\bee
   &\delta dz=dz \delta=0, \quad \deltb d\zb=d\zb \deltb=0,\\ 
   &\{\delta, d\zb\}=0, \quad \{\deltb, dz\}=0,\\
   &\delta^{2}=\deltb^{2}=0, \label{dd}\\
   &\{\delta, \deltb\}=0, \label{ddb}
\eqq
where $\{\cdot,\cdot\}$, $[\cdot,\cdot]$ are the anticommutator and commutator respectively.

\subsection{One-form Realization of  the Exterior Differential Operator $d$}

The calculus described in the previous section has a very interesting property.
There exists a one-form $\Xi$ having the property that
\be \label{dxi}
   \Xi f \mp f \Xi = \l df,
\eq
where, as usual, the minus sign applies for functions or even forms
and the plus sign for odd forms.
Indeed, it is very easy to check that
\be
   \Xi = \xi - \xi^*,
\eq
where
\be
   \xi = qdz \rho^{-1}\zb,
\eq
satisfies Eq.(\ref{dxi}) and
\be
   \Xi^* = -\Xi.
\eq
It is also easy to check that
\be
   d\Xi = 2qd\zb \rho^{-2}dz
\eq
and
\be
   \Xi^{2} = q\l d\zb \rho^{-2}dz.
\eq
Suitably normalized, $d\Xi$ is the natural area element on the quantum sphere.
Notice that $\Xi^{2}$ commutes with all functions and forms,
as required for consistency with the relation
\be
   d^{2} = 0.
\eq

The existence of the form $\Xi$ within the algebra of
$z, \zb, dz, d\zb$ is especially interesting because no such form exists
for the 3-D calculus on $SU_{q}(2)$ \cite{Wcal1},
from which we have derived the calculus on the quantum sphere
(a one-form analogous to $\Xi$ does exist for the two
bicovariant calculi on $SU_{q}(2)$,
but we have explained before why we didn't choose either of them).
It is also interesting that $d\Xi$ and $\Xi^{2}$ do not vanish
(as the corresponding expressions do in the bicovariant calculi on the quantum groups
or in the calculus on quantum Euclidean space).
The one-form $\Xi$ is regular everywhere on the sphere,
except at the point $z = \zb =\infty$,
which classically corresponds to the north pole.

\subsection{Right Invariant Vector Fields on $S^2_q$}

First let us recall some well-known facts about the vector fields on $SU_q(2)$
(see for example Ref.\cite{Z1}).
The enveloping algebra $\U$ of $SU_q(2)$ is usually said to be generated
by the left-invariant vector fields $H_{L}$,$X_{L\pm}$
which are arranged in two matrices $L^+$ and $L^-$. 
The action of these vector fields corresponds to infinitesimal
right transformation: $T\rightarrow TT'$.
What we want now is the infinitesimal version of the left transformation
given by Eq.(\ref{z-transf}), 
hence we shall use the right-invariant vector fields $H_{R}$,$X_{R\pm}$. 
Since only the right-invariant ones will be used, we will drop
the subscript ${R}$ hereafter.

The properties of the right-invariant vector fields are similar 
to those of the left-invariant ones.
Note that if an $SU_q(2)$ matrix $T$ is transformed from the right 
by another $SU_q(2)$ matrix $T'$, 
then it is equivalent to say that the $SU_{1/q}(2)$ matrix $T^{-1}$ is transformed 
from the left by another $SU_{1/q}(2)$ matrix $T'^{-1}$.
Therefore one can simply write down all properties of the left-invariant vector 
fields and then make the replacements: $q\rightarrow 1/q$, $T\rightarrow T^{-1}$ 
and left-invariant fields$\rightarrow$right-invariant fields.

Consider the matrices of vector fields:
\bee
   M^{+}= \left(
            \begin{array}{ll}
              q^{-H/2} & q^{-1/2}\l X_{+} \\
              0        & q^{H/2} 
            \end{array}
          \right),
 & 
   M^{-}= \left(
            \begin{array}{ll}
              q^{H/2}         & 0        \\
              -q^{1/2}\l X_{-} & q^{-H/2} 
            \end{array}
          \right).
\eqq
The commutation relations between the vector fields are given by,
\bee
   R_{12}M_{2}^{+}M_{1}^{+}=M_{1}^{+}M_{2}^{+}R_{12}, \\ 
   R_{12}M_{2}^{-}M_{1}^{-}=M_{1}^{-}M_{2}^{-}R_{12}, \\
   R_{12}M_{2}^{+}M_{1}^{-}=M_{1}^{-}M_{2}^{+}R_{12}, 
\eqq
while the commutation relations between the vector fields and the elements of the quantum 
matrix in the smash product \cite{Z1,SWZ1,M2,SW} of $\U$ and $SU_q(2)$ are
\bee
   T_{1}M_{2}^{+}=M_{2}^{+}{\cal R}_{12}T_{1}, \\
   T_{1}M_{2}^{-}=M_{2}^{-}{\cal R}_{21}^{-1}T_{1},
\eqq
where $T$ is an $SU_q(2)$ matrix, ${\cal R}=q^{-1/2}R$ and $R$ is the $GL_q(2)$ R-matrix.
Clearly $M^{+}$, and $M^{-}$ are the right-invariant counterparts of $L^{+}$ and $L^{-}$. 
The commutation relations between the $M$'s and the $T$'s tell us how the functions on $SU_q(2)$ are 
transformed by the vector fields $H$,$X_{+}$,$X_{-}$.
 
It is convenient to define a different basis for the vector fields,
\bee
  &\Zp=X_{+} q^{H/2}, \\
  &\Zm=q^{H/2}X_{-}
\eqq
and
\bee
   &\H=[H]_{q}=\frac{q^{2H}-1}{q^{2}-1}.
\eqq
They satisfy the commutation relations
\bee \label{ZZH}
&\H \Zp - q^4 \Zp \H =(1+q^2)\Zp, \\
    &\Zm \H - q^4 \H \Zm =(1+q^2) \Zm 
\eqq
and
\be
   q\Zp\Zm-q^{-1}\Zm\Zp=\H.
\eq
Using the expressions of $z, \zb$ in terms of $\a, \b, \g, \d$, one can easily find the action
of these vector fields on the variables $z, \zb$:
\bee
   &\Zp z=q^{2}z\Zp+q^{1/2}z^{2}, \label{vect1}\\
   &\Zp\zb=q^{-2}\zb\Zp +q^{-3/2}, \label{Zpzb} \\
   &\H z=q^{4}z\H+(1+q^{2})z, \label{Hz} \\
   &\H\zb=q^{-4}\zb\H-q^{-4}(1+q^{2})\zb, \label{Hzb} \\
   &\Zm z =q^2 z \Zm -q^{1/2} \label{Zmz}
\eqq
and
\bee
   &\Zm \zb= q^{-2} \zb \Zm -q^{-3/2} \zb^2. \label{vect2}
\eqq
It is clear that a $*$-involution can be given: 
\be \label{Z*}
\Zp^*=\Zm, \quad  \H^*=\H.
\eq

Since all the relations listed above are closed in the vector fields and $z$, $\zb$  
(this would not be the case if we had used the left-invariant fields), 
we can now take these equations as the definition of the vector fields that generate the 
fractional transformation on $S^2_q$.
We shall take our vector fields to commute with the exterior differentiation $d$.
One can show that 
this is consistent for right-invariant vector fields in a 
left-covariant calculus
and allows us to obtain the action of our vector fields on the differentials $dz$ and $d\zb$,
as well as on the derivatives $\del$ and $\delb$.
For instance (\ref{vect1}) gives
\be
   \Zp dz = q^{2}dz\Zp + q^{1/2}(dzz+zdz)
\eq
and
\be
   \del\Zp = q^{2}\Zp\del + q^{-3/2}(1+q^{2})z\del.
\eq

It is interesting to see how $\Xi$ and $d\Xi$ transform
under the action of the right invariant vector fields
or under the coaction of the fractional transformations (\ref{z-transf}).
Using (\ref{vect1}) to (\ref{vect2}) one finds
\be
   \Zp\Xi = \Xi\Zp + q^{-1/2}dz \label{ZpXi}
\eq
and
\be
   \H\Xi = \Xi\H. \label{HXi}
\eq
Eqs.(\ref{ZpXi}) and (\ref{HXi}) imply that $d\Xi$ commutes with $Z_{\pm}$ and $\H$,
as expected for the invariant area element.

For the fractional transformation (\ref{z-transf}) one finds
$\xi\rightarrow \xi'$ where
\be
   \xi'-\xi = -q(dz)cd^{-1}(1+cd^{-1}z)^{-1} \label{xi-transf}
\eq
and a similar formula for $\xi^*$.
The right hand side of (\ref{xi-transf}) is a closed one-form,
since $(dz)^{2} = 0$, so one could write
\be \label{conn}
   \xi'-\xi = -qd[\log_{q}(1+cd^{-1}z)]
\eq
with a suitably defined quantum function $\log_{q}$.
Because of Eq.(\ref{conn}), $\xi$ can be interpreted as a connection.
At any rate
\be
   d\xi' = d\xi,
\eq
so that the area element two-form is invariant under
finite transformations as well.

\subsection{The Poisson Sphere}\label{sPoisson}

The commutation relations of the previous sections give us, in the limit $q \rightarrow 1$,
a Poisson structure on the sphere. The Poisson Brackets (P.B.s) are obtained as usual as  a limit
\be \label{PB-sq2}
   (f,g) = \lim_{h \rightarrow 0} \frac{fg\mp gf}{h}, \quad q^2=e^h=1+h+[h^2],
\eq
where we use $+$ for $f,g$ both odd and $-$ otherwise.
For instance, the commutation relation (\ref{szz}) gives
\be z \zb =(1-h) \zb z -h +[h^2] \eq
and therefore \cite{We}
\footnote{
Note that this Poisson structure is not the one
usually considered on the sphere,
$(z,\zb)=\r^2$, which is associated with the symplectic structure
given by the K\"{a}hler (area) form on the sphere.}
\be (\zb, z)=\rho. \eq
Similarly one finds
\bee
&(dz,z)=zdz, &(d\zb,z)=z d\zb, \\
&(dz, \zb)=-\zb dz, &(d\zb, \zb)=-\zb d\zb
\eqq
and
\bee &(d \zb, dz)=d\zb dz. \eqq
In this classical limit functions and forms commute or anticommute according to their even 
or odd parity, as usual.  The P.B. of two even
quantities or of an even and an odd quantity is antisymmetric, that of two odd quantities is
symmetric. It is
\be d(f,g)=(df, g) \pm (f, dg), \label{Leb}\eq
where the plus (minus) sign applies for even (odd) $f$. Notice that we have enlarged the concept
of Poisson bracket to include differential forms. This is very natural when considering the classical
limit of our commutation relations.

In the classical limit, Eq.(\ref{dxi}) becomes
\be \label{pb-dxi} (\Xi, f)=df, \eq
where 
\be \Xi=\xi-\xi^* \eq
and
\be 
     \xi=dz \zb \rho^{-1}, \quad \xi^*=d\zb z \rho^{-1} 
\eq
are ordinary classical differential forms. Now
\be d\Xi=2d\zb dz \rho^{-2} \eq
and 
\be \Xi^2=0. \eq

Inspired by this example and by those of $CP_q(N)$ and the Grassmannians given in
Secs.4 and 6, it is natural to consider  the problem of constructing a
Poisson structure on the algebra of differential forms so that
(\ref{Leb}), (\ref{pb-dxi}) and some other natural conditions are valid.
This is attempted in \cite{CHZPA} and interesting results are obtained.

As before, the variables $z$ and $\zb$ cover the sphere
except for the north pole, while $w=1/z$ and $\wb=1/\zb$ miss
the south pole. It is
\be (\wb, w)=\wb w(1+\wb w). \eq
The Poisson structure is not symmetric between the north and south pole.
All P.B.s of regular functions and forms 
vanish at the north pole $w=\wb=0$.
Therefore, for Eq.(\ref{pb-dxi}) to be valid, the one-form $\Xi$
must be singular at the north pole. Indeed one finds 
\be \label{xi} \xi =\frac{dw \wb}{1+\wb w} -\frac{dw}{w}, \quad
\xi^* =\frac{d\wb w}{1+\wb w} -\frac{d\wb}{\wb}
\eq
and
\be \Xi=\frac{w d\wb-\wb dw}{\wb w (1+\wb w)}. \eq
On the other hand the area two-form
\be \label{dXi0} d \Xi =2 \frac{d\wb dw}{(1+\wb w)^2} \equiv \Omega \eq
is regular everywhere on the sphere.

The singularity of $\Xi$ at the north pole is not a real problem if we treat it in the sense  of
the theory of distributions. Consider a circle $C$ of radius $r$ encircling the origin of the $w$ plane in
a counter-clockwise direction and set
\bee &w=r e^{i\theta}, &\wb =r e^{-i \theta}. \eqq
Using (\ref{xi}), we have 
\be
\int \Xi = \int \frac {\wb dw-w d\wb}{1+\wb w}  -4 \pi i.
\eq
As $r \rightarrow 0$ the integral in the right hand side tends to zero
because the integrand is regular at the origin.
Stokes' theorem can be satisfied
even at the origin if we modify Eq.(\ref{dXi0}) to read
\be
   d\Xi= \Omega -4 \pi i \d(w) \d(\wb) d\wb dw. \label{dxi1}
\eq
It is 
\be \int_{S^2} \Omega =4 \pi i \eq
so that
\be \int_{S^2} d \Xi =0 \eq   
as it should be for a compact manifold without boundary.
Notice that the additional delta function term in (\ref{dxi1})
also has zero P.B.s with all functions and forms as required by consistency.

\subsection{Braided Quantum Spheres}
We first review the
general formulation \cite{M1} for obtaining the braiding
of quantum spaces in terms of
the universal R-matrix of the quantum group which
coacts on the quantum space. 
\subsubsection{Braiding for Quantum Group Comodules}\label{Bcomodule}
Let $\CA$ be the algebra of functions on a  quantum group and $\V$  an
algebra on which $\CA$ coacts from the left:
\bee \Delta_L: \V &\rightarrow& \CA \otimes \V \nn \\
               v & \mapsto & v^{(1')} \otimes v^{(2)}, 
\eqq
where we have used the Sweedler-like notation for $\Delta_L(v)$.

Let $\W$ be another left $\CA$-comodule algebra,
\bee
   \Delta_L: \W &\rightarrow& \CA \otimes \W \nn \\
               w & \mapsto & w^{(1')} \otimes w^{(2)}. 
\eqq
It is known \cite{M1} that one can
put $\V$ and $\W$ into a single left $\CA$-comodule algebra
with the  multiplication between
elements of $\V$ and $\W$ given by
\be \label{bcr} vw = \CR(w^{(1')},v^{(1')}) w^{(2)} v^{(2)}. \eq
Here $\CR \in \U \otimes \U$ is the universal R-matrix for the quantum
enveloping
algebra $\U$ dual to $\CA$ (with respect to the pairing $\la\cdot, \cdot\ra$)
and
\be \CR(a,b) =\la\CR, a\otimes b\ra. \eq
It satisfies:
\bee
  &\label{R0} \CR(f_{(1)}, g_{(1)}) f_{(2)} g_{(2)} = g_{(1)} f_{(1)} \CR (f_{(2)},g_{(2)}),\\
  &\label{R1} \CR(fg, h)=\CR(f,h_{(1)}) \CR(g,h_{(2)}), \\
  &\label{R2} \CR(f, gh)=\CR(f_{(1)},h) \CR(f_{(2)},g), \\
  &\label{R3} \CR(1,f) = \CR(f,1) = \epsilon(f).
\eqq

One can check that (\ref{bcr}) is associative and is left-covariant
\footnote{
If on the other hand, one starts with two right $\CA$-comodule algebras,
\bee 
\Delta_R: \V &\rightarrow& \V \otimes \CA \nn \\
               v & \mapsto & v^{(1)} \otimes v^{(2')},\\
\Delta_R: \W &\rightarrow& \W \otimes \CA \nn \\
               w & \mapsto & w^{(1)} \otimes w^{(2')}, 
\eqq
then the multiplication
\be vw = w^{(1)} v^{(1)} \CR(v^{(2')}, w^{(2')}) \eq
is associative (under the corresponding assumption), right covariant under $\Delta_R$ and
makes $\V$ and $\W$ together a right $\CA$-comodule algebra.
}.
\be
   \Delta_L(vw)= \Delta_L(v) \Delta_L(w).
\eq

For $\CA=SU_q(2)$, it is
\be \CR(T^i_j, T^k_l) = q^{-1/2} \hat{R}^{ki}_{jl},
\eq
where $\hat{R}$ is the $GL_q(2)$ R-matrix.

The braiding formula (\ref{bcr}) can be used for any number of
ordered $\CA$-comodules $\{\V_{n}\}_{n=1}^{N}$ so that it holds for
$v\in \V_m$ and $w\in \V_n$ if $m<n$.

Since we know how $z,z'$ and $\zb'$ transform, we can use (\ref{bcr}) to derive
the braided commutation relations \cite{CHZBS2}. 
We will not repeat the derivation here but will only give 
the results
\be z \zb =q^{-2} \zb z -\lambda q^{-1}, \label{z01}  \eq
\be z z' =q^2 z' z -\lambda q z'^2, \label{zz'} \eq
\be z \zb' =q^{-2} \zb' z -\lambda q^{-1}. \label{zzb'} \eq
For consistency with the $*$-involution of the braided algebra
the braiding order of $z, \zb, z'$ and $\zb'$
has to be $z<z'<\zb'<\zb$
after we have fixed $z<z'$ and $z<\zb$
as assumed in \cite{CHZS2}.
It is crucial that we braid separately $\CA=\langle\{1, z\}\rangle$ with
${\CA}'$ and $\bar{\CA}'$, and $\bar{\CA}=\langle\{1, \zb\}\rangle$ with
${\CA}'$ and $\bar{\CA}'$ instead of
simply braiding the whole algebra $\langle\{1, z, \zb\}\rangle$ with
$\langle\{1, z', \zb'\}\rangle$.
Otherwise we will not be able to have the usual properties of the $*$-involution
(e.g. $(f(z)g(z'))^*=g(z')^* f(z)^*$) for the braiding relations.

An alternative derivation of the same 
braiding relations proceeds by first computing
the braiding of two copies of the
complex quantum plane on which $SU_q(2)$ coacts
and then using the expressions 
of the stereographic variables $z$ and $\zb$ in terms of the coordinates
$x,y$ of the quantum plane
\be z=x y^{-1}, \quad \zb=\bar{y}^{-1} \bar{x}. \eq

\subsubsection{Anharmonic Ratios}\label{cross}

Let us first review the classical case.
The coordinates $x,y$ on a plane transform as
\be \label{transf-xy}
    \left(\begin{array}{c}
            x \\
            y
         \end{array}\right)\rightarrow
   \left(\begin{array}{cc}
            a & b \\
            c & d
         \end{array}\right)
   \left(\begin{array}{c}
            x \\
            y
         \end{array}\right)
\eq
by an $SL(2)$ matrix
$T=\pmatrix{a&b\cr c&d}$.
(Since here we do not need the complex conjugates $\bar{x}$ and $\bar{y}$,
$T$ does not have to be an $SU(2)$ matrix.)
The determinant-like object
$xy'-yx'$ defined for $x,y$ together with
the coordinates of another point $x',y'$
is invariant under the $SL(2)$ transformation.
For each point we define $z=x/y$ so that
\be z-z'=y^{-1}(xy'-yx')y'^{-1}. \eq
It now follows that with $x_i,y_i$ for $i=1,2,3,4$
as coordinates of four points,
\bee & &(z_2-z_1)(z_2-z_4)^{-1}(z_3-z_4)(z_3-z_1)^{-1} \nn \\
&=&(x_1 y_2-y_1 x_2)(x_4 y_2-y_4 x_2)^{-1}
(x_4 y_3-y_4 x_3)(x_1 y_3-y_1 x_3)^{-1} 
\eqq
is invariant because all the factors $y_i^{-1}$ cancel
and only the invariant parts $(x_i y_j-y_i x_j)$ survive.
Therefore the anharmonic ratio is invariant under the $SL(2)$ transformation.
(In fact it is invariant for $T$ being a $GL(2)$ matrix.)

Permuting the indices in the above expression
we may get other anharmonic ratios,
but they are all functions of the one above.
For example,
\be 
(z_2-z_3)(z_2-z_4)^{-1}(z_1-z_4)(z_3-z_1)^{-1}=
(z_2-z_1)(z_2-z_4)^{-1}(z_3-z_4)(z_3-z_1)^{-1}-1.
\eq

The coordinates of the $SU_q(2)$ covariant quantum plane obey
\be xy=qyx, \eq
an equation covariant under the transformation (\ref{transf-xy})
with $T$ now being an $SU_q(2)$ matrix.
Braided quantum planes can be introduced by using (\ref{bcr}). 
Let  $\V$ be the $i$-th copy and $\W$ be the $j$-th one, then  
we have for $i<j$,
\bee &x_i y_j =qy_j x_i +q \lambda x_j y_i,  \nn\\ 
     &x_i x_j =q^2 x_j x_i,  \nn\\
     &y_i y_j =q^2 y_j y_i, \nn\\
     &y_i x_j =q x_j y_i. 
\eqq 
In the deformed case
we have to be more careful about the ordering.
Let the deformed determinant-like object be
\be (ij)=x_i y_j-q y_i x_j, \eq
which is invariant under the $SU_q(2)$ transformation,
and let
\be [ij] = z_i - z_j = q^{-1} y_i^{-1} (ij) y_j^{-1}, \eq
where $z_i = x_i y_i^{-1}$.

Using the relations
\be y_i(ij)=q (ij)y_i, \eq
\be (ij)y_j=q y_j(ij) \eq
for $i<j$ and
\be y_i(jk)=q^3 (jk)y_i, \eq
\be (ij)y_k=q^3 y_k(ij) \eq
for $i<j<k$,
we can see that, for example,
\be A=[12][24]^{-1}[34][13]^{-1} \eq
is again invariant.
Similarly,
$B=[12][23]^{-1}[34][14]^{-1}$
as well as a number of others are invariant.

To find out whether  these invariants are
independent of one another,
we now discuss the algebra of the $[ij]$'s.
                                                                   
Because $[ij]=[ik]+[kj]$ and $[ij]=-[ji]$
the algebra of $[ij]$ for $i,j=1,2,3,4$ is generated
by only three elements $[12],[23],[34]$.
It is easy to prove that
\be
   [ij][kl]=q^{2}[kl][ij]
\eq
if $i<j\leq k<l$.

It follows that we have
\be
   [ij][ik][jk]=q^4[jk][ik][ij] \label{YB}
\eq
for $i<j<k$, and
\be
   [12][34]+[14][23]=[12][24]+[24][23].
\eq
Using  these relations we can check
the dependency between the
different anharmonic 
ratios. 
For example, let $C=[13][23]^{-1}[24][14]^{-1}$,
and $D=[14][13]^{-1}[23][24]^{-1}$, both invariant,
then
\bee
   B^{-1}AC &=& 1  \\
   q^{2}B-D^{-1} &=& -1.
\eqq

In this manner it can be checked that
all products of four terms
$[ij]$, $[kl]$, $[mn]^{-1}$, $[pr]^{-1}$
in arbitrary order, which are invariant, are  functions of
only one invariant, say, $A$.
Namely, all invariants are related and
just like in the classical case,
there is only one independent anharmonic ratio.
If one uses the $SU_q(2)$ covariant commutation relations
(\ref{z01}) and (\ref{zzb'}), one can  check that the anharmonic ratio commutes
with all the $\zb_i$'s and so commutes with its $*$-complex
conjugate, which is also an invariant. 

\subsection{Integration}\label{Integration}

We want to determine the invariant integral $\la f \ra$  of a function $f(z,\zb)$ over the sphere.

\subsubsection{Using the Definition}

A left-invariant integral can be defined,
up to a normalization constant,
by requiring invariance under the action of the right-invariant
vector fields
\be \label{xi-int}
   \la\chi f(z,\zb)\ra = 0
\eq
for $\chi = \Zp,\Zm,\H.$

Using $\H$ and Eqs.(\ref{Hz}) and (\ref{Hzb}) one finds that
\be
   \la z^{k}\zb^{l}g(\zb z)\ra = 0,\;\; \mbox{ unless} \;\; k=l.
\eq
(Here $g$ is a convergence function such that $z^{k}\zb^{l}g(\zb z)$ belongs to the sphere.)
Therefore we can restrict ourselves to integrals of the form $\la f(\zb z)\ra$.

Eqs.(\ref{vect1}) and (\ref{Zpzb}) imply
\be
   \Zp\r = \r\Zp + q^{1/2}z\r
\eq
and
\be
   \Zp\r^{-l} = \r^{-l}\Zp-q^{-3/2}[l]_{1/q}z\r^{-l}.
\eq
From $\la\Zp(\zb\r^{-l})\ra = 0$, $l\geq 1$,
one finds easily the recursion formula
\be \label{recursion}
   [l+1]_{q}\la\r^{-l}\ra = [l]_{q}\la\r^{-l+1}\ra,\;\; l\geq1,
\eq
which gives
\be
   \la\r^{-l}\ra = \frac{1}{[l+1]_{q}}\la 1\ra,\;\; l\geq 0.
\eq
Similarly
\be
   \la\frac{\zb z}{(1+\zb z)^{l}}\ra = (\frac{1}{[l]_{q}}-\frac{1}{[l+1]_{q}})\la 1\ra,
    \;\; l\geq 1.
\eq
We leave it to the reader to find the expression for
\be
   \la\frac{(\zb z)^{k}}{(1+\zb z)^{l}}\ra,\;\; l\geq k.
\eq

Notice that one can also compute the integral by using
the ``cyclic property'' of the quantum integral
\footnote{
Similar cyclic properties have been found by H. Steinacker\cite{Ste}
for integrals over higher dimensional quantum spheres in quantum
Euclidean space.}
\be
 \la f(z,\zb)g(z,\zb)\ra=\la g(z,\zb)f(q^{-2}z,q^{2}\zb)\ra,
\eq
which can be derived from the requirement of invariance under the action
(\ref{xi-int}) of vector fields or from the requirement of invariance
under finite fractional transformation.

\subsubsection{Using the Braiding}

We can also compute the left-invariant integral by requiring its
consistency with the braiding relations.

Since both $z'$ and $\zb'$ are always on the same side
of either variable of their braided copy, $z$ or $\zb$,
in the braiding order ($z<z'<\zb'<\zb$),
the integration on $z',\zb'$,
has the following property:

if
\be f(z',\zb')g(z,\zb)=\sum_{i}g_{i}(z,\zb)f_{i}(z',\zb'), \eq
then
\be \la f(z',\zb')\ra g(z,\zb)=\sum_{i}g_{i}(z,\zb)\la f_{i}(z',\zb')\ra,
\label{fg} \eq
where $\la\cdot\ra$ is the invariant integral on $S_{q}^{2}$.
However,
\be f(z',\zb')\la g(z,\zb)\ra\neq\sum_{i}\la g_{i}(z,\zb)\ra f_{i}(z',\zb'). \eq
 
The above property (\ref{fg}) can be used to derive explicit integral rules.
For example, consider the case of $f(z',\zb')=\zb'\rho'^{-n}$,
where $\rho'=1+\zb' z'$ and $g(z,\zb)=z$.
Since
\be \zb'\rho'^{-n}z = q^2 z\zb'\rho'^{-n}+
q^{1-2n}\lambda([n+1]_{q}-[n]_{q}\rho')\rho'^{-n},\quad n\geq 0, \eq
using (\ref{fg}) and $\la\zb'\rho'^{-n}\ra=0$ we get
the recursion relation:
\be [n+1]_{q}\la\rho'^{-n}\ra=[n]_{q}\la\rho'^{-(n-1)}\ra, \quad n\geq 1.
\label{rr} \eq
This agree with the first method.

\section{$CP_q(N)$ AS A COMPLEX MANIFOLD} \label{CPqN}
\setcounter{equation}{0}

\subsection{$SU_q(N+1)$ Covariant Complex Quantum Space}

For completeness, we list here the formulas we shall need to 
construct the complex projective space. 
Remember that the $SU_q(N+1)$ symmetry can be represented  
\cite{ChuZ} on the 
complex quantum space $C_q^{N+1}$ with 
coordinates $x_i, \xb^i, i=0,1,...,N$, which satisfy the relations
\be \label{xx} x_i x_j =q^{-1} \RR^{kl}_{ij} x_k x_l , \eq
\be \label{xbx} \xb^i x_j=q (\RR^{-1})^{ik}_{jl} x_k \xb^l \eq
and
\be \label{xbxb} \xb^i \xb^j=q^{-1} \RR^{ji}_{lk} \xb^k \xb^l. \eq
Here $q$ is a real number, $\RR^{kl}_{ij} $ is the $GL_q(N+1)$ $\R$-matrix
\cite{FRT} with indices running from $0$ to $N$,
and $\xb^i=x_i^*$ is the $*$-conjugate of $x_i$.
The Hermitian length 
\be \label{length} L=x_i \xb^i \eq
is real and central. 
The $\RR$-matrix satisfies the characteristic equation
\be
   (\RR-q)(\RR+q^{-1})=0.
\eq

Derivatives $D^i, \Db_i$ can be introduced (the usual symbols
$\del^a, \delb_b$ are reserved below for the derivatives on 
$CP_q(N)$ ) which satisfy
\bee &D^i x_j =\delta^i_j +q \RR^{ik}_{jl} x_k D^l,
     &D^i \xb^j =q (\RR^{-1})^{ji}_{lk} \xb^k D^l,\\
     &\Db_i \xb^j =\delta^i_j +q^{-1} (\RR^{-1})^{lj}_{ki} \xb^k \Db_l,
     &\Db_i x_j =q^{-1} \Pt^{lk}_{ji} x_k \Db_l
\eqq
and 
\be D^i D^j=q^{-1} \RR^{ji}_{lk} D^k D^l, \eq
\be D^i \Db_j =q^{-1} \Pt^{ki}_{lj} \Db_k D^l, \eq
\be \Db_i \Db_j =q^{-1} \RR^{kl}_{ij} \Db_k \Db_l. \eq 
Here we have defined
\be \Pt^{ij}_{kl}= \RR^{ji}_{lk} q^{2(i-l)} =\RR^{ji}_{lk} q^{2(k-j)}, \eq
which satisfies
\be \Pt^{ri}_{sj} (\RR^{-1})^{jk}_{il}=
    (\RR^{-1})^{ri}_{sj} \Pt^{jk}_{il}= \delta^r_l \delta^k_s
\eq
and (summing over the index $k$)
\be \Pt^{ik}_{jk}=\delta^i_j q^{2i+1}, \quad 
\Pt^{ki}_{kj}=\delta^i_j q^{2(N-i)+1}. \eq

Using
\be
   \RR^{ij}_{kl}(q^{-1})=(\RR^{-1})^{ji}_{lk}(q)
\eq
and
\be
   \RR^{ij}_{kl}=\RR^{kl}_{ij},
\eq
one can show that 
there is a symmetry of this algebra:
\bee  &q\rightarrow q^{-1}, \label{s1}\\
      &x_i\rightarrow kq^{-2i}\xb^i, \quad \xb^i\rightarrow lx_i,\\
      &D^i\rightarrow k^{-1}q^{2i}\Db_i,\quad \Db_i\rightarrow l^{-1}D^i, \label{s4} 
\eqq
where $k$ and $l$ are arbitrary constants.
Exchanging the barred and unbarred quantities in (\ref{s1})-(\ref{s4}),
we get another symmetry which is related to the inverse of this one.

Using the fact that $L$ commutes with $x_i, \xb^i$, a $*$-involution can be 
defined for $D^i$
\be (D^i)^*=-q^{-2i'}L^n \Db_i L^{-n}, \eq
where 
\be i'=N-i+1 \eq
for any real number $n$.
The $*$-involutions corresponding to different $n$'s are
related to one another by the symmetry of conjugation by $L$
\be a\rightarrow L^{m}aL^{-m}, \eq
where $a$ can be any function or derivative and
$m$ is the difference in the $n$'s.

The differentials $\xi_i=dx_i, \xib^i=(\xi_i)^*$ satisfy
\bee
   \label{xxi}  &x_i \xi_j =q \RR^{kl}_{ij} \xi_k x_l, \\
   \label{xbxi} &\xb^i \xi_j =q (\RR^{-1})^{ik}_{jl} \xi_k \xb^l
\eqq
and
\bee
   &\xi_i \xi_j = -q \RR^{kl}_{ij} \xi_k \xi_l, \\
   &\xib^i \xi_j =-q (\RR^{-1})^{ik}_{jl} \xi_k \xib^l.
\eqq

All the above relations are covariant under the right $SU_q(N+1)$ transformation
\bee \label{x-transf}
      &x_i \rightarrow x_j T^j_i, &\xb^i \rightarrow (T^{-1})^i_j \xb^j,\\
      &D^i \rightarrow (T^{-1})^i_j D^j, 
      &\Db_i \rightarrow  \Db_j q^{2i'} T^j_i q^{-2j'},\\
      &\xi_i \rightarrow \xi_j T^j_i, &\xib^i \rightarrow (T^{-1})^i_j \xib^j,
\eqq 
where $T^i_j \in SU_q(N+1)$.
\footnote{
Due to our conventions of using a right $SU_q(N+1)$ covariant quantum space here, 
(\ref{xx})-(\ref{xbxb}) are different from the left-covariant ones (\ref{qplane2}) in the case of 2-dimensions.
And  as a consequence, the equations (\ref{zz}), (\ref{zzb}) etc. below for the case of $N=1$ 
are also different from what we obtained in the last section for the sphere $S_q^2$. 
}

The holomorphic and antiholomorphic differentials
$\delta, \deltb$  
satisfy the undeformed Leibniz rule, $\delta^2=\deltb^2=0$ and
 $\deltb x_j =x_j \deltb$ etc.

\subsection{Algebra and Calculus on $CP_q(N)$}

Define for $a=1,...,N$,
\footnote{The letters $a,b,c,e$ etc. run from 1 to $N$, while
$i, j, k, l$ run from 0 to $N$.}
\be z_a=x_0^{-1} x_a, \quad \zb^a=\xb^a (\xb^0)^{-1}. \eq
It follows from  (\ref{xx}) and (\ref{xbx}) that
\be
 \label{zz} z_a z_b =q^{-1}  \R^{ce}_{ab} z_c z_e, \eq
\be \label{zzb}
 \zb^a z_b=q^{-1} (\R^{-1})^{ac}_{be} z_c \zb^e -\lambda q^{-1} \delta^a_b,
\eq
where $\R^{ac}_{be}$ is the $GL_q(N)$ $\R$-matrix with indices running
from 1 to $N$.

It follows from (\ref{xxi}) and (\ref{xbxi}) that
\be z_a dz_b =q \R^{ce}_{ab} dz_c z_e \label{zdz-cpn}, \eq
\be \zb^a dz_b =q^{-1} (\R^{-1})^{ac}_{be} dz_c \zb^e, \label{zbdz} \eq
\be  dz_a dz_b = -q \R^{ce}_{ab} dz_c dz_e \label{dzdz-cpn} \eq
and
\be d\zb^a dz_b = -q^{-1} (\R^{-1})^{ac}_{be} dz_c d\zb^e. \label{dzdzb} \eq

The derivatives $\del^a, \delb_a$ are defined by requiring 
$\delta \equiv dz_a \del^a$
and $\deltb \equiv d\zb^a\delb_a$ to be exterior differentials.
It follows from (\ref{zdz-cpn}) and (\ref{zbdz}) that
\be \label{delz-cpn} \del^a z_b = \delta^a_b +q \R^{ac}_{be} z_c \del^e, \eq 
\be \del^a \zb^b = q^{-1} (\R^{-1})^{ba}_{ec} \zb^c \del^e, \eq 
\be \delb_a z_b =q \Phi^{ec}_{ba} z_c \delb_e, \eq
\be \delb_a \zb^b = \delta^b_a +q^{-1} (\R^{-1})^{eb}_{ca} \zb^c \delb_e, \eq 
\be \label{deldel} \del^b \del^a =q^{-1} \R^{ab}_{ce} \del^e \del^c \eq
and
\be \del^a \delb_b=q \Phi^{ca}_{eb} \delb_c \del^e, \eq
where the $\Phi$ matrix is defined by 
\be \Phi^{ca}_{db} =\R^{ac}_{bd} q^{2(c-b)}=\R^{ac}_{bd} q^{2(d-a)}. \eq

Similarly as in the case of quantum spaces
the algebra of the differential calculus on $CP_{q}(N)$
has the symmetry:
\bee   &q\rightarrow q^{-1}, \\
       &z_a\rightarrow rq^{-2a}\zb^a, \quad \zb^a\rightarrow sz_a,\\
       &\del^a\rightarrow r^{-1}q^{2a}\delb_a, \quad \delb_a\rightarrow s^{-1}\del^a, 
\eqq
where $rs=q^2$.
Again we also have another symmetry by exchanging the barred
and unbarred quantities and $q\rightarrow 1/q$ in the above.

Also the $*$-involutions 
\bee &z_a^*=\zb^a, \\
     &dz_a^*=d\zb^a
\eqq
and 
\be \label{inv-del} \del^{a*}=-q^{2n-2a'}\rho^{n} \delb_a \rho^{-n}, \eq
where
\be a'=N-a+1 \eq
and \be \r=1+\sum_{a=1}^{N}z_a\zb^a, \eq
can be defined for any $n$.
Corresponding to different $n$'s they are related with
one another by the symmetry of conjugation by $\rho$
to some powers followed by a rescaling by appropriate powers of $q$.

In particular, the choice $n=N+1$ gives the $*$-involution
which has the correct classical limit of Hermitian conjugation
with the standard measure $\r^{-(N+1)}$ of $CP(N)$.

The transformation (\ref{x-transf}) induces a transformation on $CP_q(N)$
\be \label{z-transf-cpn} z_a \rightarrow 
(T^0_0 +z_b T^b_0)^{-1} (T^0_a +z_c T^c_a). \eq
One can then calculate how the differentials transform
\be \label{dz-transf-cpn} dz_a \rightarrow dz_b M^b_a, 
\quad d \zb^a \rightarrow (\Md)^a_b d \zb^b, \eq
where $M^b_a$  is a matrix of functions in $z_a$ with coefficients in 
$SU_q(N+1)$ 
and $(\Md)^a_b \equiv (M^b_a)^*$.
Since $\delta, \deltb$ are invariant, the transformation 
on the derivatives follows
\be \label{del-transf} \del^a \rightarrow (M^{-1})^a_b \del^b, \quad 
    (\del^a)^* \rightarrow (\del^b)^* ((\Md)^{-1})^b_a .
\eq 
The covariance of the $CP_q(N)$ relations under the transformation
(\ref{z-transf-cpn}), (\ref{dz-transf-cpn}) and (\ref{del-transf}) follows 
directly from the covariance of $C_q^{N+1}$.

\subsection{One-Form Realization of Exterior Differentials} \label{oneform}

Let us first recall that in Connes' non-commutative geometry \cite{Con},
the calculus is quantized using
the following operator representation for the  differentials,
\be d \omega=F \omega -(-1)^k \omega F \eq
where $\omega$ is a $k$-form and  $F$ is an operator such that
$F^*=F$ and $F^2=1$.
\footnote{The appropriate setting is a Fredholm module $(\H, F)$  where all these
relations take place in the Hilbert space $\H$.}
In the bicovariant calculus on quantum groups \cite{Wcal2}, there exists a
one-form $\eta$
with the properties $\eta^*=-\eta$, $\eta^2=0$ and
\be d f= [\eta, f]_{\pm}, \eq
where $[a,b]_{\pm}=a b \pm b a$ is the graded commutator
with plus sign only when both $a$ and $b$ are odd.
It is interesting to ask when will such a realization of differentials exist?
And what will be the properties of this special one-form?
Instead of studying the operator aspect, we will first consider these questions in
the simpler algebraic sense.

\subsubsection{A Special One-Form}

Let us first look at the example of
the $SO_q(N)$ covariant quantum space \cite{FRT,oz}.
The quantum matrix $T$ of $SO_q(N)$ satisfies in addition to
\be \R_{12} T_1 T_2 =T_1 T_2 \R_{12}, \eq
also the orthogonality relations \cite{FRT}
\be T^t g T=g, \quad T g^{-1} T^t = g^{-1}, \eq
where the numerical quantum metric matrices $g=g_{ij}$ and $g^{-1}=g^{ij}$ can be
chosen to be equal $g_{ij}=g^{ij}$.
The coordinates $x_i$ of the quantum Euclidean space satisfy the commutation relations
\be x_k x_l \R^{kl}_{ij}=q x_i x_j -\lambda \alpha L g_{ij}, \eq
where $L=x_k x_l g^{kl}=x_k x^k$ and
$\alpha=\frac{1}{1+q^{N-2}}. $
The differentials of the coordinates
$ \xi_i=dx_i $
satisfy the commutation relations
\be  x_i \xi_j = q \xi_k x_l \R^{kl}_{ij}. \eq
It can be verified that
\be \label{lx} L x_i= x_i L, \quad L dx_i =q^2 dx_i L. \eq
Hence $\eta = -q^{-1}dL L^{-1}$ satisfies 
\be \lambda df =[\eta, f]_{\pm}. \eq

Generalizing this idea, 
we have the following construction:
\begin{cn}   \label{cn1}

Let $A$ be an algebra generated by coordinates $x_i$ and 
$(\Omega(A), d)$ be a differential calculus
\footnote{
By this we mean an $A$-bimodule $\Omega(A)$ generated by $x_i, dx_i$ with
commutation relations specified such that $(d 1) =0$, graded Leibniz rule
is satisfied and $d^2=0.$}
over $A$.
If there exists an element $a \in A$, unequal nonvanishing constants $r, s$ such that
\be a x_i=r x_i a , \quad a dx_i =s dx_i a, \quad \forall i, \eq
then
\be \lambda df =[\eta, f]_{\pm} \eq
with
\be \label{eta} \eta =  \frac{\lambda}{1-s/r} da a^{-1}. \eq
The normalization constant $\lambda$ is introduced so that
$\lambda/(1-s/r)$ is well defined as $r, s, q \rightarrow 1$.
\end{cn}

It is  not hard to prove that $\eta^2=d\eta=0$.
As another example, in the 
$GL_q(N)$ quantum group \cite{FRT,SWZ1},
the algebra is generated by the elements of the 
quantum matrix $T=(T^i_j)_{i,j=1,...N}$ and the
differentials $dT^i_j$.
The quantum determinant $\Delta=det_q T$ satisfies
\be \Delta T^i_j =T^i_j \Delta, \quad \Delta  dT^i_j =q^2 dT^i_j \Delta \eq
and so in this case
\be \eta=-q^{-1}d\Delta \Delta^{-1}. \eq

\subsubsection{One-form Realization of the Exterior Differential for a  $*$-Algebra}    

In the same manner as in the construction in  section \ref{cn1}, we have the following:
\begin{cn}{\em($*$-Algebra)} \label{cn2}\\
Let $A$ be a $*$-involutive algebra  with coordinates $z_i, \zb_i$
and differentials $dz_i=\delta z_i , d \zb_i= \deltb \zb_i$ such that $\zb_i= z_i^*, d\zb_i=(dz_i)^*$.
If there exists a real element $a \in A$ and real unequal nonvanishing 
constants $r, s$ such that
\be a z_i=r z_i a , \quad a dz_i =s dz_i a, \quad \forall i, \label{a}\eq
then, as easily seen,
\be \lambda \delta f =[\eta, f]_{\pm}, \quad
    \eta =  \frac{\lambda}{1-s/r} \delta a a^{-1}, \label{da}
\eq
\be \lambda \deltb f =[\etb, f]_{\pm}, \quad
    \etb =  \frac{\lambda}{1-r/s} \deltb a a^{-1} \label{dba}
\eq
and
\be \lambda d f =[\Xi, f]_{\pm}, \quad \Xi=\eta +\etb, \label{xi-eta} \eq
where $\pm$ applies for odd/even forms $f$. 
\end{cn}
Notice that (\ref{da}) and (\ref{dba}), and therefore (\ref{a}),
imply that
\be r a \delta a = s \delta a a, \quad r \deltb a a = s a \deltb a. \eq

It can be proved that $\eta^*=-\etb$ and so  $\Xi^*=-\Xi$. It  holds that
$\eta^2=\etb^2=0$. However $\Xi^2=\eta \etb + \etb \eta=
\lambda \delta \etb = \lambda \deltb \eta$  will generally
be nonzero.
Note that
\be \lambda d \Xi = [\Xi, \Xi]_+ =2 \Xi^2. \eq
Define
\be K= \delta \etb= \deltb \eta \label{K-eta} \eq
then
\be K= \frac{1}{2} d \Xi. \label{K-Xi} \eq
It follows that $dK=0$ and $K^*=K$. Thus in the case $K \neq 0$, we will call
it a K\"{a}hler form and $K^n$
\footnote{$n$ = complex dimension of the algebra. We consider only deformations
such that the Poincar\'{e} series of the deformed algebra and its classical
counterpart match.}
will be non-zero and define a real volume element for an
integral (invariant integral if $K^n$ is invariant).
$K$ also has the very nice property of commuting with 
everything
\be K z_a =z_a K, \quad K dz_a =dz_a K. \eq
We see here an example of Connes' calculus \cite{Con} of the type $F^{2} \neq 0$
rather than $F^{2} = 0$. 

We consider a few examples of this construction.
In the case of the quantum sphere $S^2_q$, 
the element $\rho=1+ \bar{z} z$ satisfies
\be \rho z = q^2 z \rho, \quad \rho dz =dz \rho.\eq
Therefore, we obtain
\be \eta=q dz \rho^{-1} \bar{z}, \quad \etb = -q d\bar{z} \rho^{-1} z \eq
and $K$ is just the area element
\be 
K=\deltb \eta = -q^3 dz d \zb \rho^{-2}. 
\eq
One can introduce the K\"{a}hler potential $V$ defined
by
\be
K= \delta \deltb V.
\eq
It is 
\be
V= \sum_{k=1}^\infty (-1)^{k-1} \frac{q^{2k-1}}{[k]_q} \zb^k z^k.
\eq

Such a one-form representation for the differential exists on both
$C_q^{N+1}$ and $CP_q(N)$.
For $C_q^{N+1}$,
we saw in the above that
\bee &L x_i =x_i L, &L \xi_i = q^2 \xi_i L \eqq
and 
\bee 
&\eta_0= -q^{-1}\delta L L^{-1},  &\etb_0 = q \deltb L L^{-1}. \eqq
In this case, $K$ is not the K\"{a}hler form one usually assigns to $C_q^{N+1}$.
Rather, it gives $C_q^{N+1}$ the geometry of $CP_q(N)$ written in 
homogeneous coordinates.

Similar relations hold for $CP_q(N)$ in inhomogeneous coordinates.
It is 
\bee &\rho z_a =q^{-2} z_a \rho, &\rho dz_a =dz_a \rho \label{ro-zdz} \eqq
and therefore
\be \eta= -q^{-1}\delta \rho \rho^{-1}, \quad \etb= q \deltb \rho \rho^{-1}. \eq
One can then compute 
\be \label{metric}
K=\deltb \eta = dz_a g^{a \bb} d \zb^b,
\eq
where the metric $g^{a \bb}$ is
\be g^{a \bb} =q^{-1} \rho^{-2} (\rho \delta_{a b} - q^2 \zb^a z_b) 
\label{gab}\eq
with inverse $g_{\bb c}$
\be g_{\bb c} g^{c \ab} = g^{a \cb} g_{\cb b} = \delta_{a b} \eq
given by
\be g_{\bb c} =q \rho (\delta_{b c}+ \zb^b z_c). \eq
This metric is the quantum deformation of the standard Fubini-Study metric
for $CP(N)$.
It is $K=\delta \deltb V$, where the K\"{a}hler potential $V$ is
\be 
V= \sum_{k=1}^\infty (-1)^{k-1} \frac{q^{2k-1}}{[k]_q} 
\sum_{1 \leq a_1, a_2, \cdots, a_k \leq N} z_{a_k} z_{a_{k-1}} 
\cdots z_{a_1} \zb^{a_1} \cdots \zb^{a_{k-1}} \zb^{a_k}. 
\eq

Notice that under the transformation (\ref{z-transf-cpn})
\be \eta \rightarrow \eta +q f^{-1} \delta f, \quad f=T^0_0 +z_b T^b_0 \eq
and so $K$ is invariant. 
From (\ref{dz-transf-cpn}) and (\ref{metric}), it follows that
\be \label{g-transf1}
       {g^{a \bb} \rightarrow (M^{-1})^a_c  g^{c \db}  ((\Md)^{-1})^\db_\bb }, 
\eq
\be \label{g-transf2}
       {g_{\bb a} \rightarrow  (\Md)^\bb_\db  {g_{\db c}}  M^c_a}.
\eq

One can show that the following form $dv_x$ in $C_q^{N+1}$
\bee dv_x&\equiv&\Pi_{j=0}^N (\xib^j L^{-1/2}) \Pi_{i=0}^N (L^{-1/2} \xi_i) \\ 
         &=&\rho^{-(N+1)} d\zb^N \cdots d \zb^1 dz_1 \cdots dz_N \cdot
     \xib^0 (\xb^0)^{-1} (x_0)^{-1} \xi_0
\eqq
is invariant.
Using this, one can prove that
\be 
   dv_z \equiv \rho^{-(N+1)} d\zb^N \cdots d \zb^1 dz_1 \cdots dz_N \label{dv}
\eq
is invariant also and is in fact equal to $K^N$ (up to a numerical factor).
The factor $\rho^{-(N+1)}$ justifies the choice $n=N+1$ for the 
involution (\ref{inv-del}).  

\subsection{Poisson Structures on $CP(N)$}\label{Poisson}

The commutation relations in the previous sections give us,
in the limit $q\rightarrow 1$,
a Poisson structure on $CP(N)$.
The Poisson Brackets (P.B.s) are obtained as the limit
(this definition differs from (\ref{PB-sq2}) by a factor of two)
\bee &(f,g)=\lim_{h\rightarrow 0}\frac{fg \mp gf}{h},
     &q=e^h=1+h+[h^2].
\eqq
It is straightforward to find
\bee   &(z_a, z_b)=z_a z_b, \hspace{1cm} a<b, \\
       &(z_a, \zb^b)= \left\{\begin{array}{ll}
                        z_a \zb^b, & a\neq b \\
                        2(1+\sum_{c=1}^{a}z_c \zb^c), &a=b 
                      \end{array},\right.\\
       &(z_a, dz_b)=\left\{\begin{array}{ll}
                              z_a dz_b+2z_b dz_a, & a<b \\
                              2z_a dz_a, & a=b \\
                              z_a dz_b, & a>b
                          \end{array},\right. \\
       &(\zb^a, dz_b)=\left\{\begin{array}{ll}
                              -\zb^a dz_b, & a\neq b \\
                              -2\sum_{c=1}^{a}\zb^c dz_c, & a=b
                            \end{array}\right.
\eqq
and those following from the $*$-involution, which satisfies
\be (f, g)^* = (g^*, f^*). \eq

The P.B. of two differential forms $f$ and $g$
of degrees $m$ and $n$ respectively satisfies
\be (f, g) = (-1)^{mn+1} (g, f). \eq
The exterior derivatives $\delta, \deltb, d$  act on the P.B.s distributively, 
for example
\be d(f, g) = (df, g) \pm (f, dg), \eq
where the plus (minus) sign applies for even (odd) $f$.
Notice that we have extended the concept of Poisson Bracket to include
differential forms.

The Fubini-Study K\"{a}hler form
\be
   K=dz_a g^{a\bb}d\zb^b
\eq
has vanishing Poisson bracket with all functions and forms
and, naturally, it is closed.

\subsection{Integration}

We now turn to the discussion of integration on $CP_q(N)$. We shall use the
notation $\la f(z,\zb)\ra$ for
the right-invariant integral of a function $f(z,\zb)$ over $CP_q(N).$
It is defined, up to a normalization factor, by requiring
\be \label{inv-int} \la\O f(z, \zb)\ra=0 \eq
for any left-invariant vector field $\O$ of $SU_q(N+1)$.
We can work out the integral by looking at the explicit action of
the vector fields on functions. This approach has been worked out for
the case of the sphere but it gets rather complicated for the higher
dimensional projective spaces.
We shall follow a different and simpler
approach here.
First we notice that the identification
\be x_i /L^{1/2}=T^N_i, \quad \xb^i /L^{1/2}=(T^{-1})^i_N,
\quad i={0,1,...,N}, \eq
where $T$ is an $SU_q(N+1)$ matrix,
reproduces (\ref{xx})-(\ref{length}). Thus if we define
\be \label{int} \la f(z,\zb)\ra \equiv \la f(z, \zb) |_{z_a=(T^N_0)^{-1} T^N_a,
    \zb^a=(T^{-1})^a_N /(T^{-1})^0_N}\ra_{SU_q(N+1)}, \eq
where $\la \cdot \ra_{SU_q(N+1)}$ is the Haar measure \cite{W1} on $SU_q(N+1)$,
then it follows immediately that (\ref{inv-int}) is satisfied.
\footnote{A similar strategy of using the ``angular'' measure to
define an integration has been employed by
H. Steinacker \cite{Ste} in constructing integration over quantum Euclidean space.}
Next we claim that
\be
\label{int1} \la (z_1)^{i_1} ( \zb^1)^{j_1} \cdots (z_N)^{i_N} (\zb^N)^{j_N}\ra =0
\;{\;  \rm unless \;}\;
    i_1=j_1,..., i_N=j_N.
\eq
This is  because the integral is invariant under the finite transformation
(\ref{z-transf-cpn}).
For the particular choice $T^i_j =\delta^i_j \alpha_i$, with
$|\alpha_i|=1, \Pi_{i=0}^N \alpha_i =1$, this gives
\be z_a \rightarrow (\alpha_a /\alpha_0) z_a \eq
and so (\ref{int1}) follows.

In \cite{W1}, Woronowicz proved the following interesting property
for the Haar measure
\be \label{int2} \la f(T) g(T)\ra_{SU_q(N+1)}= \la g(T) f(DTD) \ra_{SU_q(N+1)}, \eq
where
\be (DTD)^i_j =D^i_k T^k_m D^m_j \eq
and
\be D^i_j = q^{-N+2i} \delta^i_j \eq
is the $D$-matrix for $SU_q(N+1)$.
It follows from (\ref{int2}) that
\be
\la f(z, \zb) g(z, \zb)\ra = \la g(z, \zb) f(\D z, \D^{-1} \zb)\ra,
\label{107}
\eq
where
\be \D^a_b =\delta^a_b q^{2a}, \quad a,b =1,2,...,N .\eq

Introducing
\bee &\rho_r= 1+\sum_{a=1}^r z_a \zb^a, \eqq
one finds
\be \label{rhoz}  \rho_r z_a = \left\{ \begin{array}{ll}
                          z_a \rho_r & r<a \\
                          q^{-2}  z_a \rho_r & r \geq a
                          \end{array}
                 \right. ,
\eq
\be \rho_r \rho_s =\rho_s \rho_r \eq
and
\be \zb^a z_a= q^{-2} \rho_a - \rho_{a-1} \quad \mbox{(no sum).} \label{roro}
\eq

Because of (\ref{int1}), it is sufficient to determine integrals of the form
\be
   \la{\rho_1}^{-i_1} \cdots {\rho_N}^{-i_N} \ra. \label{ro-int} 
\eq
The values of the integers $i_a$ for (\ref{ro-int}) to make sense
will be determined later.

Consider
\bee \la\zb_a {\rho_1}^{-i_1} \cdots {\rho_N}^{-i_N}  z_a \ra&=&
        \la{\rho_1}^{-i_1} \cdots {\rho_N}^{-i_N}  z_a (q^{-2a} \zb^a)\ra\nonumber\\
     &=& q^{-2a} \la{\rho_1}^{-i_1} \cdots {\rho_N}^{-i_N}  (\rho_a-\rho_{a-1})\ra,
\eqq
where (\ref{107}) is used.
Applying (\ref{rhoz})
\bee \mbox{L.S. } 
&=&q^{2(i_a+\cdots + i_N)} \la{\rho_1}^{-i_1} \cdots {\rho_N}^{-i_N} \zb^a z_a \ra
                \nonumber\\
                &=& q^{2I_a} \la{\rho_1}^{-i_1} \cdots {\rho_N}^{-i_N} \zb^a z_a \ra,
\eqq
where we have denoted
\be I_a =i_a + \cdots + i_N. \eq
Using (\ref{roro})
we get the recursion formula
\bee 
&\la{\rho_1}^{-i_1} \cdots {\rho_{a-1}}^{-i_{a-1}+1} {\rho_a}^{-i_a} 
\cdots {\rho_N}^{-i_N} \ra [I_a+a]_q \nonumber \\
&=\la{\rho_1}^{-i_1} \cdots {\rho_{a-1}}^{-i_{a-1}} {\rho_a}^{-i_a+1} 
  \cdots {\rho_N}^{-i_N} \ra [I_a+a-1]_q.
\eqq
It is obvious then that
\be
    \la{\rho_1}^{-i_1} \cdots {\rho_a}^{-i_a} \ra
    =  \la{\rho_1}^{-i_1} \cdots {\rho_{a-1}}^{-i_{a-1}-i_a} \ra 
                                           \frac{[a]_q}{[I_a+a]_q}.
\eq
By repeated use of the recursion formula, 
$\la{\rho_1}^{-i_1} \cdots {\rho_N}^{-i_N} \ra$ reduces finally to
$\la{\rho_1}^{-i_1 -i_2 \cdots  -i_N}\ra$ 
and 
\be \la{\rho_1}^{-I_1}\ra =  \frac{1}{[I_1+1]_q} \la 1\ra. \eq

Therefore
\be 
   \la{\rho_1}^{-i_1} \cdots {\rho_N}^{-i_N} \ra=\la 1\ra\Pi_{a=1}^N \frac{[a]_q}{[I_a+a]_q}.
\eq
For this to be positive definite, $i_a$ should be restricted such that
$I_a +a > 0$ for $a=1,\cdots,N$.

\subsection{Braided $CP_{q}(N)$} \label{BraidedCPN}
As described  in \cite{CHZBS2} and also in section \ref{Bcomodule}, 
it is sufficient to know the transformation property of the algebra to derive the
braiding.
But as demonstrated there,
it is already quite complicated to obtain explicit formulas
in the case of a one dimensional algebra.
Therefore although we can derive the braiding for the $CP_{q}(N)$ using the 
general framework of \ref{Bcomodule}, we will follow a different, easier path:
first introduce the braiding for $C_q^{N+1}$ quantum planes and then use it to derive
a braiding for $CP_{q}(N)$ expressed in terms of inhomogeneous coordinates.
\subsubsection{Braided $C_q^{N+1}$}
Let the first copy of quantum plane be denoted by $x_i, \xb^i$
and the second by $x'_i, \xb'^i$ and let their 
commutation relations be:
\bee &x_i x'_j = \tau\RR_{ij}^{kl} x'_k x_l, \label{braid-xx}\\
     &\xb^i x'_j = \nu(\RR^{-1})_{jl}^{ik} x'_k \xb^l \label{braid-xxb}
\eqq
and their $*$-involutions for arbitrary numbers $\tau, \nu$.
These are consistent and covariant, as one can easily check.
One can choose $\tau=\nu^{-1}$ and  the Hermitian length $L$
will be central, $Lf'=f'L$, for any function $f'$ of $x', \xb'$.
However, $L'$ does not commute with $x, \xb$.
In the following, we don't need to assume that $\tau=\nu^{-1}$.

By assuming that the exterior derivatives of the two copies
satisfy the Leibniz rule
\bee \delta'f=\pm f\delta',
        & \deltb'f=\pm f\deltb', \\
     \delta f'=\pm f'\delta,
        & \deltb f'=\pm f'\deltb,
\eqq
where the plus (minus) signs apply for even (odd) $f$ and $f'$,
and
\bee \delta\delta'=-\delta'\delta, & \delta\deltb'=-\deltb'\delta, \\
     \deltb\delta'=-\delta'\deltb, & \deltb\deltb'=-\deltb'\deltb,
\eqq
one can derive the commutation relations between functions and
forms.
Identifying $\delta=dx_i D^i, \deltb=d\xb^i\Db_i$
for both copies, one can derive also  the
commutation relations between derivatives and functions
of different copies.
We will not write them down here.

\subsubsection{Braided $CP_{q}(N)$}
Using (\ref{braid-xx}), (\ref{braid-xxb}), one can  derive the braiding relations of
two braided copies of $CP_{q}(N)$ in terms of the inhomogeneous
coordinates
\bee   &z_a z'_b = q\R_{ab}^{ce} (z'_c -q^{-1}\lambda z_c)z_e,
        \label{zz-braid}\\
       &\zb'^a z_b = q^{-1}(\R^{-1})_{be}^{ac} z_c \zb'^e
        - q^{-1}\lambda\delta^a_b \label{zzb-braid}
\eqq
and their $*$-involutions.
Notice that these are independent of the particular choice of
$\tau$ and $\nu$.
Similarly, one can work out the commutation relations
between functions and forms of different copies
following the assumption that their exterior derivatives anticommute.
We will not list them here.

\section{QUANTUM PROJECTIVE GEOMETRY}\label{ProjGeom}
\setcounter{equation}{0}

We will show in this section that many concepts of projective geometry
have an analogue in the deformed case.
We shall study the collinearity conditions in Sec.\ref{coll-cond},
the deformed anharmonic ratios (cross ratios) in Sec.\ref{Anh},
the coplanarity conditions in Sec.\ref{copl-cond}.
In Sec.\ref{other-inv} we will show that the anharmonic ratios
are the building blocks of other invariants.

\subsection{Collinearity Condition}\label{coll-cond}
 
Classically the collinearity conditions for $m$ distinct points in $CP(N)$
can be given in terms of the inhomogeneous coordinates
$\{z^A_a| A=1,2,\cdots,m; a=1,2,\cdots,N\}$ as
\be
   (z^A_a-z^B_a)(z^C_a-z^D_a)^{-1}=(z^A_b-z^B_b)(z^C_b-z^D_b)^{-1},
   \label{ccc}
\eq
where  $A\neq B, C\neq D=1,\cdots,m$ and $a,b=1,\cdots,N$.

In the deformed case, the coordinates $\{z^A_a\}$ of $m$ points must
be braided for the commutation relations to be covariant, namely,
\be
   z^A_a z^B_b=q\R_{ab}^{ce}(z^B_c - q^{-1}\lambda z^A_c)z^A_e,\quad A\leq B,
   \label{ab-braid}
\eq
as an extension of (\ref{zz-braid}).
Eq.(\ref{zzb-braid}) can also be generalized in the same way,
but we shall not need it in this section.
This braiding has the interesting property that
the algebra of $CP_{q}(N)$ is {\em self-braided}, that is,
(\ref{ab-braid}) allows the choice $A=B$.
This property makes it possible to talk about the coincidence of points.
Actually, the whole differential calculus for braided $CP_q(N)$
described in Sec.\ref{BraidedCPN} has this property.

Another interesting fact about this braiding is that for a fixed index
$a$ the commutation relation is identical to that for braided $S_q^2$
\be
   z^A_a z^B_a=q^2 z^B_a z^A_a-q\lambda z^A_a z^A_a, \quad A\leq B.
   \label{sq2zz}
\eq

Since there is no algebraic way to say that two ``points'' are distinct
in the deformed case,
the collinearity conditions should avoid using expressions like
$(z^A_a-z^B_a)^{-1}$,
which are ill defined.
Denote
\be [AB]_a = z^A_a - z^B_a. \eq
The collinearity conditions in the deformed case can be formulated as
\be
   [AB]_a [CD]_b = q^2 [CD]_a [AB]_b, \quad \forall a,b
   \label{collinear}
\eq 
and $A<B\leq C<D$.
By (\ref{ab-braid}) this equation is formally equivalent to
the quantum counterpart of (\ref{ccc}):
\be
   [AB]_a [CD]_a^{-1} = [AB]_b [CD]_b^{-1},
   \label{collinear2}
\eq
where the ordering of $A,B,C,D$ is arbitrary.
The advantage of this formulation is that (\ref{collinear}) is a
quadratic polynomial condition and polynomials are well defined in the
braided algebra.

Therefore the algebra $Q$ of functions of $m$ collinear points
is the quotient of the algebra $\CA$ of $m$ braided copies of $CP_q(N)$
over the ideal $I=\{f\alpha g: \forall f,g\in\CA; \forall \alpha\in CC\}$
generated by $\alpha$ which stands for the collinearity conditions
(\ref{collinear}), i.e.,
$\alpha\in CC=\{[AB]_a [CD]_b - q^2 [CD]_a [AB]_b: A<B\leq C<D\}$.

Two requirements have to be checked for this definition $Q=\CA/I$ to make sense.
The first one is that for any $f\in\CA$ and $\alpha\in CC$,
\be
   f\alpha = \sum_i \alpha_i f_i, \quad \forall f\in\CA, \label{f-alpha}
\eq
for some $f_i \in\CA$ and $\alpha_i\in CC$.
This condition ensures that the ideal $I$ generated by
the collinearity conditions is not ``larger'' than what we want,
as compared with the classical case.

The second requirement is the invariance of $I$ under
the fractional transformation (\ref{z-transf-cpn}).
It can be checked that both requirements are satisfied.

\subsection{Anharmonic Ratios}\label{Anh}

Classically the anharmonic ratio of four collinear points is an invariant
of the projective mappings, which are the linear transformations
of the homogeneous coordinates.
In the deformed case, the homogeneous coordinates are the coordinates $x_i$
of the $GL_q(N+1)$-covariant quantum space,
and the linear transformations are the $GL_q(N+1)$ transformations
which induce the fractional transformations
(\ref{z-transf-cpn}) on the coordinates $z_a$ of the projective space
$CP_q(N)$.

We consider the following anharmonic ratio of $CP_q(N)$ for four collinear points
$\{z^A_a| A=1,2,3,4\}$
\be
   [A1]_a [A4]_a^{-1} [B4]_a [B1]_a^{-1}, \label{cross-ratio}
\eq
where $A,B = 2,3$.
We wish to show that it is invariant.
After some calculations and denoting $\tau(A)=[1A]_a [14]_a^{-1}$,
which is independent of the index $a$ according to the collinearity condition,
we get
\be
   [AB]_a\rightarrow U(B)^{-1}(\tau(A)-\tau(B))P_a(A)V(A)^{-1},
\eq
where
\bee
&U(B)=T^0_0+z^B_eT^e_0, \\
&V(A)=T^0_0+q z^A_f T^f_0, \\
&P_a(A)=-[14]_b M_a^b(A)
\eqq
and
\be
M_a^b(A)=(T^b_a T^0_0-q^{-1}T^b_0 T^0_a)
   + qz^A_c (T^b_a T^c_0-q^{-1}T^b_0 T^c_a).
\eq
Then the anharmonic ratio (\ref{cross-ratio}) transforms as
\bee
   [A1]_a [A4]_a^{-1} [B4]_a [B1]_a^{-1} &\rightarrow&
   U(1)^{-1}\tau(A)(1-\tau(A))^{-1}(1-\tau(B))\tau(B)^{-1}U(1)\nonumber\\
   &=& \tau(A)(1-\tau(A))^{-1}(1-\tau(B))\tau(B)^{-1} \nonumber\\
   &=& [A1]_a [A4]_a^{-1} [B4]_a [B1]_a^{-1},
\eqq
where we have used $z^1_a\tau(A) = \tau(A)z^1_a$ for any $A\geq 1$,
which is true because we can represent $\tau(A)$ as $[1A]_a [14]_a^{-1}$
with the same index $a$ and then use $z^1_a [AB]_a = q^2 [AB]_a z^1_a$.

Because of the nice property (\ref{sq2zz}), we can use the results
about the anharmonic ratios of $S_q^2$ ( which is a special case of $CP_q(N)$
with $N=1$ but no collinearity condition is needed there) in Sec.\ref{cross}.
Note that all the invariants as functions of $z^A_a$ for a fixed $a$
in $CP_q(N)$ are also invariants as functions of $z^A=z^A_a$ in $S_q^2$.
The reason is the following.
Consider the matrix $T^a_b$ defined by
\bee
   T^0_0=\alpha, & T^0_a=\beta, \\
   T^a_0=\gamma, & T^a_a=\delta, 
\eqq
where $\alpha,\beta,\gamma,\delta$ are components of an $SU_q(2)$-matrix,
$T^b_b=1$ for all $b\neq 0, a$ and
all other components vanishing.
It is a $GL_q(N+1)$-matrix, but the transformation (\ref{z-transf-cpn})
of $z^A_a$ by this matrix is the fractional transformation (\ref{z-transf})
on $S_q^2$ with coordinate $z^A=z^A_a$.

Therefore, by simply dropping the subscript $a$,
the anharmonic ratio (\ref{cross-ratio}) becomes an anharmonic ratio
of $S_q^2$.
On the other hand, since all other anharmonic ratios of $S_q^2$
are functions of only one of them,
their corresponding anharmonic ratios of $CP_q(N)$
(by putting in the subscript $a$)
would be functions of (\ref{cross-ratio})
and hence are invariant.
Therefore we have established the fact that all invariant
anharmonic ratios of $CP_q(N)$ are functions of only one of them.

\subsection{Coplanarity Condition}\label{copl-cond}

In this subsection we will get
the coplanarity condition
as a generalization of the collinearity condition
(\ref{collinear}).

For $r+1$ points spanning an $r$-dimensional hyperplane,
we have
\be
   z^B=\sum_{A\in I}\s^B_A z^A,
\eq
where $\s^B_A=(x^B_0)^{-1}\nu^B_A x^A_0$
and $\sum_{A\in I}\s^B_A = 1$.
By a change of variables for $\s^B_A$,
and letting $I=\{1,2,\cdots,r,r+1\}$, $B=0$, it is
\be
   \label{new-z}
   [01]_i=\sum_{j=1}^r\tau_j [j(j+1)]_i,
\eq
where $[AB]_i=z^A_i-z^B_i$ and
the $\tau$'s are independent linear combinations of the $\s$'s.

Choose a set $K$ of $r$ different integers from $1,2,\cdots,N$.
Consider the $r$ equations (\ref{new-z})
for $i\in K$.
Let $K=\{\a_1,\a_2,\cdots,\a_r\}$,
$M^j_i=[j(j+1)]_{\a_i}$ and $M^0_i=[01]_{\a_i}$.
Then
\be
   \label{tau}
   \tau_j=M^0_i (M^{-1})^i_j, \quad j=1,2,\cdots,r,
\eq
where $M^{-1}$ is the inverse matrix of $(M^i_j)_{i,j=1}^{r}$.

Even though $M$ is not a $GL_q(r)$-matrix we define
\be
   det_q(M)=\eps_{i_1\cdots i_r}M^1_{i_1}\cdots M^r_{i_r},
\eq
where
\be
   \eps_{\s_1\s_2\cdots\s_r}=(-q)^{l(\s)}
\eq
for $\s$ being a permutation of $r$ objects
with length $l(\s)$ and the $\eps$ tensor is $0$ otherwise.
$M^{-1}$ is then found to be
\be
   (M^{-1})^i_j=(-1)^{j-1}\eps_{i i_2\cdots i_r}M^{1}_{i_2}\cdots
   M^{j-1}_{i_j} M^{j+1}_{i_{j+1}}\cdots M^r_{i_r}(det_q(M))^{-1}.
\eq

Hence by (\ref{tau})
\be
   (-1)^{j-1}\tau_j=det_q(M(j))(det_q(M(0)))^{-1},
\eq
where
\be
   det_q(M(j))=\eps_{i_1\cdots i_r}M^0_{i_1}M^{1}_{i_2}\cdots
   M^{j-1}_{i_j} M^{j+1}_{i_{j+1}}\cdots M^r_{i_r}
\eq
(so that $det_q(M(0))=det_q(M)$).

Since this solution of $\tau$ is
independent of the choice of $K$,
by choosing another set $K'$ we have
another matrix $M'$ and
$(-1)^{j-1}\tau_j=det_q(M'(j))(det_q(M'(0)))^{-1}$.
Therefore we get the coplanarity condition
\be
\label{coplanar}
   det_q(M(j))(det_q(M(0)))^{-1}=det_q(M'(j))(det_q(M'(0)))^{-1}
\eq
for all $j=1,\cdots,r$ and any two sets of indices $K$ and $K'$.
This is obviously equivalent to
\be
   det_q(M(j))(det_q(M(k)))^{-1}=det_q(M'(j))(det_q(M'(k)))^{-1}
\eq
for all $j,k=0,\cdots,r$.

If $N\geq 2r$ then one can choose $K<K'$,
i.e., any element in $K$ is smaller than any element in $K'$,
then one can show that
\be
   det_q(M(0))det_q(M'(0))=q^r det_q(M'(0))det_q(M(0))
\eq
and a polynomial type of coplanarity condition is available:
\be \label{pol-cop}
   det_q(M(j))det_q(M'(k))=q^r det_q(M'(j))det_q(M(k)).
\eq
The algebra of functions of $r+1$ coplanar points is then
the quotient of the algebra generated by $\{z^A\}_{A=0}^{r}$
over the ideal generated by (\ref{pol-cop}).

\subsection{Other Invariants}\label{other-inv}

The anharmonic ratios are important because they are the building blocks
of invariants in classical projective geometry.
For example, in the $N$-dimensional classical case for given $2(N+1)$ points
with homogeneous coordinates $\{x^A_i\}$, inhomogeneous coordinates $\{z^A_a\}$
where $A=1,\cdots,2(N+1)$, $i=0,1,\cdots,N$ and $a=1,\cdots,N$,
we can construct an invariant
\be
   I=\frac{det(x^1,x^2,\cdots,x^N,x^{N+1})det(x^{N+2},x^{N+3},\cdots,x^{2(N+1)})}
           {det(x^1,x^2,\cdots,x^N,x^{N+2})det(x^{N+1},x^{N+3},\cdots,x^{2(N+1)})},
   \label{I}
\eq
where $det(x^{A_0},\cdots,x^{A_N})$ is the determinant of the matrix
$M^i_j=x^{A_i}_j$, $i,j=0,\cdots,N$,
which equals the determinant of the matrix
\be
   \left(\begin{array}{ccc}
             1         & \cdots & 1         \\
             z^{A_0}_1 & \cdots & z^{A_N}_1 \\
             \vdots    & \ddots & \vdots    \\
             z^{A_0}_N & \cdots & z^{A_N}_N
         \end{array}\right)
\eq
multiplied by the factor $x^{A_0}_0\cdots x^{A_N}_0$,
which cancels between the numerator and denominator of $I$.
It can be shown that this invariant $I$ is in fact the anharmonic ratio
of four points $z,z',z^{N+1},z^{N+2}$, where $z$ ($z'$) is the intersection
of the line fixed by $z^{N+1},z^{N+2}$ with the $(N-1)$-dimensional subspace
fixed by $z^1,\cdots,z^N$ ($z^{N+3},\cdots,z^{2(N+1)}$).

For the case of $N=2$ (see Fig.4.1),
$I$ is the ratio of the areas of four triangles:
\be
  I=\frac{\Triangle_{123}}{\Triangle_{124}}
     \frac{\Triangle_{456}}{\Triangle_{356}},
\eq
which is easily found to be
\be
  I=\frac{\overline{A3}}{\overline{A4}}\frac{\overline{B4}}{\overline{B3}},
\eq
the anharmonic ratio of the four points $A,B,3,4$.

\begin{figure}
   \epsfxsize=4.5in
   \hspace{0.5in}
   \epsfbox{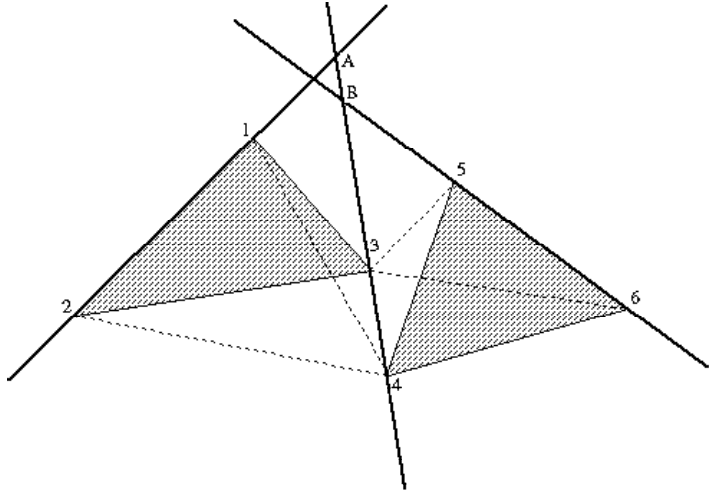}
   \vspace{-3in}
   \caption{The invariant $I$ as a cross ratio of $A,B,3,4$.}
\end{figure}

It is remarkable that all this can also be done in the quantum case.
One can construct an invariant $I_q$ using the quantum determinant
and describe the intersection between subspaces
of arbitrary dimension spanned by given points.
It is shown that the invariant $I_q$ is indeed an anharmonic ratio
in the same sense as the classical case.

\section{QUANTUM GRASSMANNIANS $G^{M,N}_q$} \label{Grass}
\setcounter{equation}{0}
In this section, we study the quantum deformation of the Grassmannians.
\subsection{The Algebra}
Let $C^i_a$, $i=1, 2, \cdots, M$, $a=1,2, \cdots, M+N$, be an $M \times (M+N)$ 
rectangular matrix satisfying the commutation relations
\be  \R^{'ij}_{ kl} C^k_c C^l_d =C^i_a C^j_b {\R}^{ab}_{cd}, \eq 
where $\R'^{ij}_{kl}$ is a $GL_q(M)$ $\R$-matrix, with indices $i,j,k,l$ etc. 
 going from 1 to $M$ and
 $\R^{ab}_{cd}$ is a $GL_q(M+N)$ $\R$-matrix, with indices $a,b,c,d$ etc. 
going from 1 to $M+N$.
In compact notation, it is
\be \label{RCC}
\R'_{12} C_1 C_2 =C_1 C_2 \R_{12}
\eq
and
(\ref{RCC}) is right-covariant under the transformation
\be \label{T1} C \rightarrow CT, \eq
where $T^a_b$ is a $GL_q(M+N)$ quantum matrix
and is also left-covariant  under  the transformation
\be \label{T2} C \rightarrow SC, \eq
where $S^i_j$ is a $GL_q(M)$ quantum matrix.
Writing
\be C^i_a =(A^i_j, B^i_\a) 
\eq
with $\a =1, 2, \cdots, N$,
we have
\bee \label{RAB}
\R'_{12} A_1 A_2 &=& A_1 A_2 \R'_{12}, \nn \\
\R'_{12} B_1 B_2 &=& B_1 B_2 \R''_{12}, \nn \\
A_1 B_2 &=& R'_{21} B_2 A_1,
\eqq
where $\R^{''\a\b}_{  \g\d}$ is a $GL_q(N)$ $\R$-matrix, with indices $\a,\b,\g,\d$ etc.
 going from 1 to $N$.

Define the coordinates $Z^i_\a$ for the quantum Grassmannians 
$G^{M,N}_q$
\be Z=A^{-1} B. \eq
$Z$ is invariant under the transformation (\ref{T2}), while under (\ref{T1}),
it transforms as
\be Z\rightarrow (\a +Z \g)^{-1} (\b +Z \d), 
\eq
where $\a,\b,\g,\d$ are the sub-matrices of $T$
\be 
T=\pmatrix{\a&\b \cr \g&\d}.
\eq
It follows from (\ref{RAB}) that $Z$ satisfies
\be 
\R'_{21} Z_1 Z_2 =Z_1 Z_2 \R''_{12}.
\eq

\subsubsection{$*-$structure}
We consider $q$ to be a real number. One can introduce the $*-$conjugate
variables $(C^i_a)^*$ and impose the commutation relation
\be C_1^\dagger \R'^{-1}_{12} C_1 =C_2 \R^{-1}_{12} C_2^\dagger, \eq
i.e.
\bee 
(A^{-1})^\dagger_2 \R'_{12} (A^{-1})_2 &=& (A^{-1})_1 \R'_{12}  (A^{-1})^\dagger_1, \nn \\
B^\dagger_1 A_2^{-1} &=& A_2^{-1} B^\dagger_1 R'^{-1}_{12}, \nn \\
B^\dagger_1 \R^{'-1}_{12} B_1 &=& B_2 \R^{''-1}_{12} B^\dagger_2 - \lambda I_1 (A A^\dagger)_2,
\eqq
where $(I_1)^{\a}_{\b} = \d ^\a_\b$ is the identity matrix.
These imply
\be
Z^\dagger_1 \R'_{21} Z_1 = Z_2 \R''^{-1}_{12} Z_2^\dagger - 
\lambda I_1 I_2. 
\eq
Explicitly,
\be 
(Z^\dagger)^\a_i \R^{'si}_{tj} Z^j_\b = Z^s_\g (\R^{'' -1})^{\a \g}_{\b \d} (Z^\dagger)^\d_t
- \lambda \d^\a_\b \d^s_t. 
\eq

\subsection{Calculus}
One can introduce the following commutation relation for functions and one-forms
\be \R^{' -1}_{12} C_1 dC_2 =dC_1 C_2 \R_{12},
\eq 
i.e.
\bee 
\R^{'-1}_{12} A_1 dA_2 &=& dA_1 A_2 \R'_{12}, \nn \\
\R^{'-1}_{12} B_1 dB_2 &=& dB_1 B_2 \R''_{12}, \nn \\
dA_1 B_2 &=& R'^{-1}_{12} B_2 dA_1, \nn \\
A_1 dB_2 &=& R'_{21} (dB_2 A_1 + \lambda dA_2 B_1 P_{12}),
\eqq
where $(P_{12})^{ij}_{kl} = \delta^i_l \delta^j_k.$
Since $Z^i_\a= (A^{-1})^i_k B^k_\a$, it is easy to derive
\be dZ = A^{-1}(dB-dA Z) \eq
and
\bee \label{dZA}
Z_1 dA_2 &=& dA_2 R^{'-1}_{12} Z_1, \nn \\
dZ_1 A_2 &=& A_2 R^{'-1}_{12} dZ_1, \nn \\
Z_1 dB_2 &=& (dB_2 Z_1-\lambda Z_1 (dA Z)_2 P_{12}) R''_{12}.
\eqq
It follows
\be \R^{'-1}_{21} Z_1 dZ_2 =dZ_1 Z_2 \R''_{12}.
\eq

To introduce a $*-$structure for the calculus, it is consistent to take
$(dZ^i_\a)^*=d(Z^{i *}_\a)$.
In addition we impose a complex structure on the calculus so that
$d=\delta+\bar{\delta}$, where $\delta$ ($\bar{\delta}$) acts only
on the holomorphic (antiholomorphic) part, satisfies Eqs.(\ref{dd})
and (\ref{ddb}).
This implies, after some calculation,
\be 
Z^\dagger_1 \R'_{21} dZ_1 = dZ_2 \R^{''-1}_{12} Z_2^\dagger. 
\eq

\subsection{One-Form Realization}
Introduce the matrix
\be E^i_j =C^i_a (C^{\dagger})^a_j.
\eq
It is 
\be
E^\dagger = E \eq
and 
\be \label{EE}
\R^{'-1}_{12}E_1 \R^{'-1}_{12} E_1 = E_1 \R^{'-1}_{12} E_1 \R^{'-1}_{12}.
\eq
One can show that
\be R_{I, J}^{'-1} E_J R_{J, I}^{'-1} E_I = E_I R_{I, J}^{'-1} E_J R_{J, I}^{'-1}
\eq
or equivalently,
\be  R'_{I, J}  E_I \bullet E_J =E_J \bullet E_I R'_{I, J}. \eq
The bullet product is defined \cite{SWZ2} inductively by
\be E_I \bullet E_J \equiv E_I R_{I, J}^{'-1} E_J R'_{I, J} \eq
for any $I=(1' 2' \cdots m')$, $J=(1 2 \cdots n),$ 
where
\bee E_{(1 2 \cdots M)} &\equiv& E_1 \bullet E_2 \bullet 
\cdots \bullet E_M \nn\\
&=&E_1(R^{'-1}_{12} E_2 R'_{12}) \cdots 
(R^{'-1}_{(M-1) M}\cdots R^{'-1}_{1 M} E_M R'_{1M} R'_{2M} \cdots R'_{(M-1)M}),
\eqq
\be \label{BigR}
\begin{array}{rcl}
R_{I,I\!I} &\equiv& \la\CR, A_I \tens A_{I\!I}\ra  \\
  &   =  &
           \begin{array}[t]{cccc}
           R_{1'n}&\cdot \; R_{1'(n-1)}&\cdot \;\ldots\;\cdot & R_{1'1}\\
  \cdot \; R_{2'n}&\cdot \; R_{2'(n-1)}&\cdot \;\ldots\;\cdot & R_{2'1}\\
        \vdots    &    \vdots          &                    &\vdots  \\
  \cdot \; R_{m'n}&\cdot \; R_{m'(n-1)}&\cdot \;\ldots\;\cdot & R_{m'1},
           \end{array}
  \end{array}
\eq
$\CR$ is the universal R-matrix for $GL_q(M)$
and
\be \label{AI}
A_I  \equiv A_{(1' 2' \cdots m')} \equiv A_{1'} A_{2'}\cdots A_{m'},
\quad
A_{I\!I} \equiv A_{(1 2 \cdots n)}  \equiv A_{1} A_{2}\cdots A_{n}.
\eq
Hence one can introduce the quantum determinant \cite{SWZ2} for the generators $E$,
\be Det E \eps^{1 2 \cdots M}  = E_{(1 2 \cdots M)} \eps^{1 2 \cdots M}, \eq
where
$\eps^{1 2 \cdots M}$ is the $\eps$ tensor  for $GL_q(M)$.

Denote
\be L= Det E \eq
and one can show that 
\bee
L E^i_j &=& E^i_j L, \\
C L &=& L C
\eqq
and
\be
dC L = q^{-2} L dC.
\eq
Using the general procedure stated in Sec.\ref{oneform} , we obtain the realization on the algebra generated by $C^i_a, dC^i_a$ and their $*-$conjugates,
\be 
\eta = -qL^{-1} \delta L. 
\eq

To find the one-form realization for the exterior differential operating on
the complex Grassmannians $Z, dZ$, we introduce
\bee
X^k_l &=& (A^{-1})^k_i E^i_j (A^{\dagger -1})^j_l \nn\\
      &=& \d^k_l +Z^k_\a (Z^\dagger)^\a_l.
\eqq
It is not hard to check that
\be
\R'_{12} X_2 \R'_{12} X_2 = X_2 \R'_{12} X_2 \R'_{12}.
\eq
Since $X$ commutes like the vector field $Y$, the quantum determinant
\bee 
\rho &\equiv& Det X \nn\\
&=& X_{(1 2 \cdots M)} \eps^{1 2 \cdots M}
\eqq
is central in the algebra of $X$.
Here, the $\bullet$-product for $X$ is
\be X_I \bullet X_J = R^{'-1}_{I J} X_I R'_{IJ} X_J, 
\eq
as for the vector field $Y$.
In particular,
\bee
X_{(1 2 \cdots M)}&=&
(R^{'-1}_{12} R^{'-1}_{23} \ldots R^{'-1}_{1M} X_1 R_{1M} \ldots R_{12})
\cdot 
(R^{'-1}_{23}R^{'-1}_{24} \ldots R^{'-1}_{2M} X_2 R_{2M} \ldots R_{23})\nn\\
&&
\cdots\cdot 
(R^{'-1}_{(M-1)M} X_{M-1} R_{(M-1)M}) X_M.
\eqq  

Introducing the quantum determinants $det (A^{-1})$, $det (A^{\dagger -1})$ 
\bee 
det (A^{-1}) \eps^{1 2 \cdots M} &=&  A_M^{-1} \cdots A_2^{-1} A_1^{-1}  \eps^{1 2 \cdots M},\nn\\
det (A^{\dagger -1}) \eps^{1 2 \cdots M} &=&  
(A^{\dagger -1})_1  (A^{\dagger -1})_2 \cdots (A^{\dagger -1})_M \eps^{1 2 \cdots M} 
\eqq
for $A^{-1}$ and $A^{\dagger -1}$ satisfying the ``RTT''-like relations
\bee
\R'_{12}(q^{-1}) A_1^{-1} A_2^{-1} &=& A_1^{-1} A_2^{-1} \R'_{12}(q^{-1}),\nn\\
\R'_{12} (A^{\dagger -1})_1 (A^{\dagger -1})_2 &=&  
(A^{\dagger -1})_1 (A^{\dagger -1})_2 \R'_{12},
\eqq
it can be shown that
\bee
&\rho = det (A^{-1}) \, L \,det (A^{\dagger -1}), \\
&Z \rho = q^2 \rho Z
\eqq
and
\be
dZ \rho = \rho dZ.
\eq
As a result, we have the one-form realization
\be
\eta =-q^{-1} \rho^{-1} \delta \rho
\eq
for the exterior derivative acting on
the algebra generated by $Z^i_\a,dZ^i_\a$
and their $*$-conjugates. The  K\"{a}hler form
\be \label{KGr} 
K=\deltb \eta
\eq
is central as usual.

\subsection{Braided $G^{M,N}_q$}
Let $Z, Z'$ be two copies of the quantum Grassmannians $G^{M,N}_q$
defined by
\be Z=A^{-1} B, \quad Z'=A^{' -1} B',
\eq
where $C^i_a =(A^i_j, B^i_\a)$, $C^{'i}_a =(A^{'i}_j, B^{'i}_\a)$
both satisfy the relations (\ref{RCC}).
Let the mixed commutation relations be
\be \label{RCC'}
Q_{12} C_1 C'_2 =C'_1 C_2 \R_{12},
\eq
where $Q$ is a numerical matrix.
For (\ref{RCC'}) to be consistent with (\ref{RCC}),
we can take  $Q$ to be $\R'^{\pm 1}$.
For either of these two choice, (\ref{RCC'}) is
covariant under
\be C \rightarrow CT, \quad C' \rightarrow C'T, \eq
where $T^a_b$ is a $GL_q(M+N)$ quantum matrix
and also under  the transformation
\be C \rightarrow SC, \quad C' \rightarrow SC', \eq
where $S^i_j$ is a $GL_q(M)$ quantum matrix.
We will pick $Q=\R'$ in the following
\be
\R'_{12} C_1 C'_2 =C'_1 C_2 \R_{12}.
\eq
Explicitly, it is 
\bee \label{RAB'}
R'_{12} A_1 A'_2 &=& A'_2 A_1 R'_{12}, \nn \\
R'_{12} B_1 B'_2 &=& B'_2 B_1 R''_{12}, \nn \\
B_1 A'_2 &=& R^{'-1}_{12} A'_2 B_1, \nn \\
A_1 B'_2 &=& R^{'-1}_{12} B'_2 A_1 +\lambda B_1 A'_2 P_{12}.
\eqq
It follows that \footnote{
If we had made the other choice $Q= \R^{' -1}$ in the above, 
the relations (\ref{RAB'}) would be different, but  
(\ref{ZZ'}) would remain the same.}
\be \label{ZZ'}
Z_1 Z'_2 = R'_{12} Z'_2 Z_1 R''_{12} -\lambda Z_1 Z_2 \R''_{12}.
\eq

One can introduce a $*$-structure to this braided algebra, the
relation 
\be
C_1^\dagger \R^{'-1}_{12} C'_1 = C'_2 \R^{-1}_{12} C_2^\dagger
\eq
is consistent and is covariant under 
\be C \rightarrow CT, \quad C' \rightarrow C'T, \eq
and
\be C \rightarrow SC, \quad  C' \rightarrow SC' \eq
with the same $T, S$ quantum matrices as explained before.
It follows immediately
\be \label{Z*Z'}
Z^\dagger_1 \R'_{21} Z'_1 = Z'_2 \R^{''-1}_{12} Z^\dagger_2 -\lambda I_{1}I_{2}.
\eq
One can also show that the K\"{a}hler form $K$
of the original copy (\ref{KGr}) 
commutes also with the $Z',Z^{' \dagger}, dZ', dZ^{' \dagger}$.

This concludes our discussion for the quantum Grassmannians,
with the case of complex projective spaces 
$CP_q(N)=G^{1,N}_q$ 
as a special case.\footnote{
Notice that for $M=1$, the numerical R-matrix becomes a number:  $\R'_{12} =q.$} 


\section{Acknowledgement}

This work was supported in part by the Director, Office of
Energy Research, Office of High Energy and Nuclear Physics, Division of
High Energy Physics of the U.S. Department of Energy under Contract
DE-AC03-76SF00098 and in part by the National Science Foundation under
grant PHY-9514797.

\appendix
               
\section{RELATION TO CONNES' FORMULATION} \label{Connes}

\renewcommand{\theequation}{\Alph{section}.\arabic{equation}}
\setcounter{equation}{0}

Here we make a comment on the relation of our work to Connes' 
quantum Riemannian geometry \cite{Con}.
We will try to re-formulate
the differential and integral calculus on the quantum sphere
in a way as close to his formulation as possible.
We take $0<q\leq 1$.

To do so we first give the spectral triple $(X, {\cal H}, \Dir)$ for this case.
$X$ is the algebra of functions on $S_q^2$.
${\cal H}$ is the Hilbert space on which both functions and
differential forms are realized as operators.
It is chosen to be composed of two parts
${\cal H}={\cal H}_0 \otimes V$.
The first part ${\cal H}_0$ is any Hilbert space
representing the algebra X.
An example is \cite{P1}
\bee
   \pi(z)|n\ra   &=& (q^{-2n}-1)^{1/2}|n-1\ra, \nn \\
   \pi(\zb)|n\ra &=& (q^{-2(n+1)}-1)^{1/2}|n+1\ra, \quad n=0,1,2,\cdots.
   \label{irrep}
\eqq
Another example is the Gel'fand-Na\v{i}mark-Siegel construction
using the integration $\la~\cdot~\ra$ introduced in Sec.\ref{Integration}.
The second part $V$ is $\C^2$, as in the classical case.
Operators on ${\cal H}$ are therefore $2\times 2$ matrices
with entries being operators on ${\cal H}_0$.
Finally, the Dirac operator is an anti-self-adjoint operator
\footnote{It is a pure convention that we choose $\Dir$
to be anti-self-adjoint rather than self-adjoint
like Connes usually does.}
on ${\cal H}$:
\be
   \Dir = k\left(\begin{array}{cc}
                   i & \pi(\zb) \\
                   -\pi(z) & -i
                \end{array}\right),
\eq
where $k$ is a real number.

According to Connes we proceed as follows to find
the differential calculus on $S_q^2$.
The representation $\pi$ of $X$ on ${\cal H}_0$ is extended
to be a representation on ${\cal H}$
for the universal differential calculus $\Omega_X$ by
\be
   \pi(a_0(da_1)\cdots(da_n))=\pi(a_0)[\Dir,\pi(a_1)]\cdots[\Dir,\pi(a_n)].
\eq
In particular, one finds
\bee
   \pi(dz)|\psi\ra
      &=& q^{-1}\l k\pi(\r)\tau|\psi\ra, \\
   \pi(d\zb)|\psi\ra
      &=& q^{-1}\l k\pi(\r)\tau^{\dagger}|\psi\ra, \quad |\psi\ra\in{\cal H},
\eqq
where the $\gamma$-matrices
      $\tau = \left(
                 \begin{array}{ll}
                     0 & 1 \\
                     0 & 0
                 \end{array}
              \right)$,
$\tau^{\dagger} = \left(
                 \begin{array}{ll}
                     0 & 0 \\
                     1 & 0
                 \end{array}
                 \right)$
satisfy the deformed Clifford algebra
$q\tau\tau^{\dagger}+q^{-1}\tau^{\dagger}\tau = \I$ for
        $\I = \left(
                 \begin{array}{ll}
                     q & 0 \\
                     0 & q^{-1}
                 \end{array}
              \right)$.

It can be checked that the kernel of the map $\pi$
for one-forms in $\Omega_X$, is generated from
\be
   z(dz)-q^{-2}(dz)z, \quad z(d\zb)-q^{-2}(d\zb)z,
\eq
and their $*$-involutions, by multiplying with functions from both sides.
The kernel of $\pi$ for two-forms is generated from
\be
   (dz)(dz)=0, \quad (d\zb)(d\zb)=0
\eq
by multiplying with functions and from the kernel of one-forms
by multiplying with one-froms from both sides.
The auxiliary fields form the ideal $Aux$ defined to be the sum
of the kernels of all degrees and the differential of them.
So in our case $Aux$, in addition to the sum of kernels
mentioned above, is generated by
\be
   d[z(d\zb)-q^{-2}(d\zb)z]=(dz)(d\zb)+q^{-2}(d\zb)(dz).
\eq
The other differentials are already contained in the sum.
According to Connes,
the differential calculus is obtained from the spectral triple by
\be
   \Omega(X)=\Omega_X/Aux.
\eq
This gives precisely the same covariant differential calculus
described in Sec.\ref{DC-sq2}.

Next we consider the integration on $S_q^2$.
Connes' formula for the integration is
\be \label{Connes-int}
   \int\a=Tr_{\om}(\gamma\pi(\a)|\Dir|^{-d}),
\eq
where $Tr_{\om}$ is the Dixmier's trace \cite{Dix}
and $\gamma$ is the $\Z_2$-grading operator.
Here $d$ is the dimension of the quantum space,
which, according to Connes, is defined by
the series of eigenvalues of $|\Dir|^{-1}$.
In our case $d$ determined that way is zero.

One should expect that Connes'
prescription will not give the same invariant integration
on $S_q^2$ (\ref{xi-int}) because while
Connes' integration always has the cyclic property
\be
   \int\om_1\om_2=\pm\int\om_2\om_1,
\eq
we know that the $SU_q(2)$-invariant integration does not.
Remarkably, if we choose to use the classical
dimension $d=2$ of the two-sphere
in the formula (\ref{Connes-int}),
we actually obtain the invariant integration.
This is shown in the following.

Note that the calculus on $S_q^2$ is $\Z_{2}$-graded by
\be
   \g = k^{-2}\pi((dz d\zb-d\zb dz)\r^{-2})
      = \left(\begin{array}{cc}   
            1 & 0 \\
            0 & -1
        \end{array}\right),
\eq
which satisfies
\bee
   &\g^2=I, \quad \g^{\dagger}=\g, \\
   &\g\pi(a)=\pi(a)\g \quad \forall a\in X, \\
   &\g \Dir = -\Dir \g.
\eqq

We define the integration on $S_q^2$ by the trace
\be \label{con-int}
   \int \a = Tr(\g\pi(\a) |\Dir|^{-2}),
\eq
where $Tr$ is the appropriate trace on the Hilbert space ${\cal H}$.
(If the Hilbert space of (\ref{irrep}) is used for ${\cal H}_0$,
one should simply use the ordinary trace.)
It can be directly checked that the integration is compatible
with the differential calculus
\be
    \int Aux = 0, \label{aux}
\eq
by using the representations of the auxiliary fields
\be
   \pi(a)\left(\begin{array}{cc}   
                q & 0 \\
                0 & q^{-1}
            \end{array}\right)
\eq
for any $a\in X$ (including $0$) and
\be
   |\Dir|^{-2} = qk^{-2}\pi(\r^{-1})
                \left(\begin{array}{cc}
                    q^{-1} & 0 \\
                    0 & q
                \end{array}\right).
\eq
Using the cyclic property of the trace,
it follows also that Stokes' theorem
\be
   \int d\a = 0 \label{Stokes}
\eq
is valid for any one-form $\a$.

Since Stokes' theorem can be used
to derive the recursion relations (\ref{recursion}),
the integration (\ref{con-int}) coincides with
the invariant integration on $S_q^2$ up to normalization.

Note that Eqs.(\ref{aux}) and (\ref{Stokes}) are valid
for any choice of ${\cal H}_0$,
hence the formula (\ref{con-int}) gives the same
invariant integration as long as an appropriate trace exists
so that our selected integrable functions, e.g. $\r^{-n}$ for $n\geq 0$,
multiplied by the area element (K\"{a}hler two-form) have
finite integrals.

\bibliographystyle{srt}

\end{document}